\documentclass[onecolumn,showkeys,preprintnumbers,amsmath,amssymb]{revtex4}

\usepackage{rotating}
\usepackage{makecell}
\usepackage{array,multirow}
\usepackage{bbm}

\usepackage[normalem]{ulem}
\usepackage{graphicx,epsfig}
\usepackage{dcolumn}
\usepackage{bm}
\usepackage{xcolor}

\usepackage{orcidlink}

\newcommand{\req}[1]{Eq.~(\ref{#1})}

\newcommand{\fig}[1]{Fig.~\ref{#1}}
\newcommand{\tab}[1]{Table \ref{#1}}

\usepackage{multirow}
\usepackage{xcolor}

\newcommand{\red}{\textcolor{red}}
\newcommand{\blue}{\textcolor{blue}}

\newcommand{\cut}[1]{{}}

\makeatletter
\newcommand{\thickhline}{%
    \noalign {\ifnum 0=`}\fi \hrule height 1pt
    \futurelet \reserved@a \@xhline
}
\newcolumntype{"}{@{\hskip\tabcolsep\vrule width 2pt\hskip\tabcolsep}}
\makeatother

\newcommand{\beginsupplement}{%
        \setcounter{table}{0}
        \renewcommand{\thetable}{S\arabic{table}}%
        \setcounter{figure}{0}
        \renewcommand{\thefigure}{S\arabic{figure}}%
     }

\NewDocumentEnvironment{alignb}{b}{%
	\begin{align*}
		\refstepcounter{equation} #1 \tag{\theequation}
	\end{align*}
}{}

\begin{document}


\title{Inferring Structure of Cortical Neuronal Networks from Firing Data: A Statistical Physics Approach}
\author{
Ho Fai Po{\Large \,\orcidlink{0000-0002-3056-4064}}$^1$\footnote{poh@aston.ac.uk}, 
Akke Mats Houben{\Large \,\orcidlink{0000-0002-8215-7882}}$^{2,3}$, 
Anna-Christina Haeb{\Large \,\orcidlink{0009-0008-0194-1036}}$^{2,3}$, \\ 
David Rhys Jenkins{\Large \,\orcidlink{0009-0008-7107-0838}}$^4$,  
Eric J. Hill{\Large \,\orcidlink{0000-0002-9419-1500}}$^5$, 
H. Rheinallt Parri{\Large \,\orcidlink{0000-0002-1412-2688}}$^4$, 
Jordi Soriano{\Large \,\orcidlink{0000-0003-2676-815X}}$^{2,3}$, 
David Saad{\Large \,\orcidlink{0000-0001-9821-2623}}$^1$
}

\affiliation{$^1$Department of Mathematics, Aston University, Birmingham B4 7ET, United Kingdom \\ $^2$Departament de Física de la Matèria Condensada, Universitat de Barcelona, Barcelona, Spain\\ $^3$ Universitat de Barcelona Institute of Complex Systems(UBICS), 08028 Barcelona, Spain \\ $^4$ College of Health and Life Sciences, Aston University, Birmingham B4 7ET, United Kingdom \\ $^5$Department of Chemistry, Loughborough University, Loughborough, Leicestershire LE11 3TU, United Kingdom}
\date{\today}

\begin{abstract}
Understanding the relation between cortical neuronal network structure and neuronal activity is a fundamental unresolved question in neuroscience, with implications to our understanding of the mechanism by which neuronal networks evolve over time, spontaneously or under stimulation. It requires a method for inferring the structure and composition of a network from neuronal activities. Tracking the evolution of networks and their changing functionality will provide invaluable insight into the occurrence of plasticity and the underlying learning process. We devise a probabilistic method for inferring the effective network structure by integrating techniques from Bayesian statistics, statistical physics and principled machine learning. The method and resulting algorithm allow one to infer the effective network structure, identify the excitatory and inhibitory nature of its constituents, and predict neuronal spiking activities by employing the inferred structure. We validate the method and algorithm's performance using synthetic data, spontaneous activity of an \emph{in silico} emulator and realistic \emph{in vitro} neuronal networks of modular and homogeneous connectivity, demonstrating excellent structure inference and activity prediction. We also show that our method outperforms commonly used existing methods for inferring neuronal network structure. Inferring the evolving effective structure of neuronal networks will provide new insight into the learning process due to stimulation in general and will facilitate the development of neuron-based circuits with computing capabilities. 
\end{abstract}


\maketitle


\section{Introduction}

Rvealing how cortical neuronal networks connectivity evolves in time, spontaneously or under stimulation is a foundational question in neuroscience, in particular, understanding how cortical neurons learn from repeated stimulation through changes in topology and synaptic strengths~\cite{maeda1995mechanisms, orlandi2013noise, eytan2006dynamics, wagenaar2006extremely, cohen2008determinants, pasquale2008self, tetzlaff2010self}. While there are many tools for investigating macroscopic brain activities and changes, such as fMRI (functional magnetic resonance imaging), MEG (magnetoencephalography) and EEG (electroencephalography) \cite{buzsaki2004neuronal, berger1929elektroenkephalogramm, logothetis2008we}, investigating the microscopic changes which occur in neuronal tissues non-invasively remains a challenge. While in-vivo interrogation of neuronal networks at the microscopic level remains difficult, recent techniques such as MEA (microelectrode array)~\cite{maccione2010experimental} facilitate the monitoring and stimulation of neuronal tissues at cellular resolution and open the way to greater understanding of the learning process. New developments in the application of neuron-based circuits with computing capabilities make the need to understand the exact relationship between learning and stimulation more urgent and relevant~\cite{sumi2023biological, moriya2019quantitative, zanini2023investigating, parodi2023deepening}.

Machine learning (ML) and artificial intelligence (AI) play an increasingly crucial role in our daily life. However, training ML and AI systems require unsustainable computing power and energy consumption~\cite{strubell2019energy} and mostly lack the ability to adapt their structure in response to a changing situation. This has given rise to the search for alternative computing paradigms, in particular, the emerging field of biological computation, which aims at employing human neuronal networks (hNNs) as processing units in biological computing devices. These developments, such as using cortical brain organoids for non-linear curve prediction~\cite{cai2023brain} and employing cortical neuronal networks for decision-making in simulated gaming environments~\cite{kagan2022vitro}, have recently drawn the attention of both researchers and the general public. These breakthroughs point to the immense potential of hNNs in biological machine learning. The common belief is that plasticity and learning occur through appropriate stimulation in neuronal-based computing devices, such that both network topology and synaptic strengths evolve to structures that can carry out specific data/stimulation-driven tasks. However, the tools needed to investigate the evolving structure and support the understanding of how these hNNs-based devices operate are currently lacking. Hence, it is crucial to develop a principled inference tool that can reveal the effective cortical network structure, to better comprehend the mechanism that gives rise to task learning from stimulation.

Various methods have been employed to infer the effective neuronal network from its firing patterns. Commonly used techniques include generalized transfer entropy (GTE)~\cite{orlandi2014transfer}, dynamic causal modeling and Granger causality~\cite{friston2011functional}. However, these methods have significant limitations. They can only measure directional causation between neurons and identify the existence of an effective connection by setting an appropriate but somewhat arbitrary threshold. These methods cannot find the excitatory or inhibitory nature of neurons without manipulating the network through stimulation or channel blocking~\cite{orlandi2014transfer, ito2011extending}, nor can they determine the model's effective synaptic strengths. As such, they are less suitable for studies requiring long-term monitoring and careful consideration of stimulation protocols as they may potentially affect network development as observed in the training of cortical neurons-based learning machines~\cite{cai2023brain,kagan2022vitro}.

Moreover, these methods often overlook the activities of nearby neurons and fail to capture interactions between multiple neurons, resulting in inaccurate inference. Additionally, methods such as GTE do not provide a probabilistic model for neuronal activities, making it difficult to predict or reproduce network activities using the inferred effective connectivity structure for validation or further investigations. 

To fill these gaps, here we advocate mapping neuronal activities onto the kinetic Ising model of statistical physics as they share some common features, such as binary state of activity, unidirectional non-equilibrium and nonlinear dynamics and multi-neuron interactions. Mapping neuronal activities onto the kinetic Ising model facilitates the inference of inter-neuron interactions~\cite{ terada2018objective, terada2020inferring}. Yet, inferring the kinetic Ising model structure and properties is challenging and probabilistic methods have been developed in the statistical physics community for inferring the underlying directional interaction strengths from observation sequences~\cite{roudi2011mean, mezard2011exact, lombardi2023statistical}. However, these methods have been derived for a simple model, where coupling strengths are Gaussian distributed with mean zero and small variance, which does not hold in biological neuronal networks. Thus, a principled probabilistic method that can identify connectivity, synaptic strengths, and the excitatory/inhibitory nature of each neuron is required.

In this paper, we introduce an algorithm that combines models from statistical physics, Bayesian inference and probabilistic machine learning to infer the effective architecture of biological neuronal networks from firing patterns. Our proposed algorithm overcomes some of the limitations of existing methods and infers not only the effective connectivity from neuronal firing but also the neuronal characteristics (inhibitory/excitatory) and the existence of connections. Furthermore, unlike conventional methods, our algorithm provides a probabilistic model that facilitates the simulation of neuronal activities using the inferred architecture, which can be used for structure validation, prediction and further investigations. We evaluate the performance of our algorithm using synthetically generated data, \emph{in silico} neuronal network emulator data, and calcium imaging recordings of real \emph{in vitro} cortical networks with patterned~\cite{montala2022rich} and unpatterned substrates, demonstrating excellent agreement between the effective inferred model and the corresponding data.

\section{Model}
\textit{Kinetic Ising model.---} The kinetic Ising model~\cite{fredrickson1984kinetic} in statistical physics studies the activity of spins in a system with asymmetric coupling strengths. Sharing the common feature that spin (neuron) configuration is influenced by directional coupling (synaptic) strengths and the state of its neighbors, the kinetic Ising model is suitable for describing neuronal spiking activities. Here, we map the binary neuronal spiking activity onto the kinetic Ising model, which is a discrete-time non-equilibrium probabilistic structure. Consider a system of $N$ neurons within an interacting neuronal network. We denote a discrete variable $s_i^t =\pm1$ when neuron $i$ is spiking or silent at time step $t$, respectively, for $i=1,\dots,N$. Previous works~\cite{orlandi2014transfer} suggest that considering interaction across multiple time intervals have minimal effect on the inference, so we define the transitional probability of neuron $i$ at time $t$, given the neuronal activities at time $t-1$, as
\begin{align}
	P\left(s_{i}^{t}\left|\boldsymbol{s}^{t-1},\boldsymbol{J},H_{i}\right.\right)=\frac{\exp\left[\left(H_{i}+\underset{j}{\sum}J_{ij}s_{j}^{t-1}\right)s_{i}^{t}\right]}{2\cosh\left(H_{i}+\sum_{j}J_{ij}s_{j}^{t-1}\right)}, \label{eq_trans_prob}
\end{align}
where $\boldsymbol{J}\!=\!\left \{ J_{ij} \right \}_{ij}$ and $J_{ij}$ denotes the synaptic strength from $j$ to $i$, and $H_i$ denotes the external local field acting on neuron $i$, which can be interpreted as the activeness of $i$ when no signal is received from its neighbors. A positive (negative) $J_{ij}$ represents the excitatory (inhibitory) strength of a signal sent from neuron $j$ to neuron $i$ when $j$ spikes, while $J_{ij}\!=\!0$ indicates that neuron $j$ is not effectively connected to $i$. Since the activities of neurons at time $t$ depend only on activities in the previous time-step and hence do not exhibit explicit same-bin inter-dependence, the probability of activities for all neurons is given by $P\left(\boldsymbol{s}^{t}\left|\boldsymbol{s}^{t-1},\boldsymbol{J},\boldsymbol{H}\right.\right)=\prod_{i}P\left(s_{i}^{t}\left|\boldsymbol{s}^{t-1},\boldsymbol{J},H_{i}\right.\right)$. In order to infer the neuronal nature and effective link existence, we introduce two sets of latent variables, $z_j=\pm 1$ representing the excitatory and inhibitory nature of neuron $j$, respectively; and $\phi_{ij} = \{1,0\}$ indicating whether $j$ is connected to $i$ or not, respectively. One of the limitations of the kinetic Ising model is that it does not consider longer synaptic time delays, but as indicated in a number of studies~\cite{orlandi2013}, their influence is much weaker than that of single time step delays.

\textit{Adapting Ising model to living neuronal networks.---}  Based on the mathematical model defined, we introduce prior distributions to align with real-world neuronal network properties. For clarity, we use $p(\cdot)$ to denote prior distributions for the corresponding variables, while $P(\cdot)$ denotes the probability more generally. We define
\begin{align}
	p\left(z_{j}\right)&=\gamma\delta_{z_{j},+1} + \left(1-\gamma\right) \delta_{z_{j},-1}; \label{eq_prior_z_j} \\
	p\left(\phi_{ij} \left| a \right. \right)&= \delta_{\phi_{ij}, 1} \theta_{ij}e^{-al_{ij}} + \delta_{\phi_{ij}, 0}\left(1- \theta_{ij}e^{-al_{ij}}  \right); \label{eq_prior_phi_ij} \\
	p\left(J_{ij}\right)&=\sum_{z_{j}}\sum_{\phi_{ij}}p\left(J_{ij}\left|\phi_{ij},z_{j}\right.\right)p\left(\phi_{ij}\right)p\left(z_{j}\right); \label{eq_prior_J_j}\\
	p\left(H_{i}\right)&=\frac{e^{-\frac{\left(H_{i}-\mu_{H}\right)^{2}}{2v_{H}}}}{\sqrt{2\pi v_{H}}}, \label{eq_prior_H_i}
\end{align}
where $\gamma \in [0,1]$ represents the proportion of excitatory neurons in the network; $\theta_{ij}$ is the prior probability for the existence of a link from neuron $j$ to neuron $i$, which can be adjusted for convergence assistance in networks with predefined or dictated connectivity~\cite{montala2022rich,yamamoto2023modular}, or set to 1 otherwise; $a\in \mathbb{R^+}$ is a controlling parameter that accounts for the decay in connection probability; and $l_{ij}=l_{ji}$ the Euclidean distance between $i$ and $j$, thus $e^{-al_{ij}}$ reflects the exponential decay in connection probability with distance;  $\mu_H$ and $v_H \in \mathbb{R}$ represent the mean and variance of the distribution for $\boldsymbol{H}$, reflecting the distribution of neuron inherent activeness. The distribution of $J_{ij}$ comprises a mixture of distributions since it is conditioned on whether there exists a link between two neurons and whether the nature of the interaction is inhibitory or excitatory; the different cases may be characterized by different parameters. We define the conditional probabilities $p\left(J_{ij}\left|0,z_j\right.\right) = \delta\left(J_{ij}\right)\approx{\cal N}\left(0,\epsilon\right)$ for small $\epsilon$ (disconnected case) and $p\left(J_{ij}\left|1, z_j \right.\right)=\mathbbm{1}_{z_j J_{ij}>0} ~e^{-\left(\ln z_j J_{ij}-\mu_{J}^{z_j}\right)^{2}/\left(2v_{J}^{z_j}\right)}/\sqrt{2\pi v_{J}^{z_j}}$ for $\mu_J^{z_j} , v_J^{z_j} \in \mathbb{R^+}$ and $ z_j = \pm 1$, where $\mathbbm{1}$ is an indicator function ensuring all connections from $j$ are of the same sign, and $z_j J_{ij} $ follows a log-normal distribution to ensure the probability of coupling strengths of inhibitory or excitatory neurons sum to $1$. The compact notation merely deals with the positive (excitatory) and negative (inhibitory) interaction strengths within the same expression by allowing for different distribution characteristics. For brevity, we denote $\boldsymbol{\rho}=\left\{ \gamma,a,\mu_{H},v_{H},\mu_{J}^{\pm},v_{J}^{\pm}\right\} $ as the set of all defined hyperparameters (parameters that control the distributions of $\boldsymbol{J}$ and $\boldsymbol{H}$). By combining the distributions we define the evidence function as
\begin{align}
	\ln P\left(\mathbf{s}\left|\boldsymbol{J},\boldsymbol{H},\boldsymbol{\rho}\right.\right)+\ln p\left(\boldsymbol{H}\left|\boldsymbol{\rho}\right.\right)+\ln p\left(\boldsymbol{J}\left|\boldsymbol{\rho}\right.\right).
	\label{eq_evidence}  
\end{align}
In principle, the related probability of Eq.~(\ref{eq_evidence}) should be integrated over $\boldsymbol{J}$ and $\boldsymbol{H}$ to obtain the marginalized probability $P\left(\mathbf{s}\left|\boldsymbol{\rho}\right.\right)$ but is approximated at the peak of the posterior (`Maximum A Posteriori', MAP)~\cite{bishop2006pattern}.

\textit{Connectivity network inference.--- }
We extend the kinetic Ising model by introducing two sets of latent variables and prior distributions for all variables to align with the nature of realistic biological neuronal networks. The Methods section and the Supporting Information (SI) provide details on how the (latent) variables and hyperparameters can be evaluated in a principled way, enabling us to interpret the neuronal type from $z_j$, link existence through $\phi_{ij}$, and effective synaptic strength from $J_{ij}$. In general, we are applying the generalized maximum likelihood (GML) approach (whereby parameters and hyperparameters are iteratively determined, also termed the evidence procedure) and the variational expectation maximization (EM) algorithm~\cite{bishop2006pattern} to infer $\boldsymbol{J}$, $\boldsymbol{H}$ and $\boldsymbol{\rho}$. 

\section{Results}
To validate the efficacy of our model and inference algorithm, we performed tests using synthetic data, data generated by \emph{in silico} models and data from biological, \emph{in vitro} cortical neuronal activity recordings. 

In the SI, we show that the proposed method performs much better compared to the mean field and maximum likelihood approaches. In our GML inference for the kinetic Ising model, coupling strengths are not zero-mean and of small variance, and more importantly, the method allows one to infer the existence of connections between neurons. Here, we focus on the results obtained from the \emph{in silico} model with patterned substrates as well as the two realistic \emph{in vitro} cortical neuronal activities. One of these in vitro activities is of a homogeneous network with potassium stimulation, while the other focuses on spontaneous activities in a neuronal network with patterned substrates.

\subsubsection{\emph{in silico} experimental data}
\begin{figure*}
	\centering
	\begin{tabular}{@{} c @{} c @{} }
		\begin{tabular}{l}
			A \\
			\includegraphics[width=0.2375\linewidth, trim=0 0 0 0, clip]{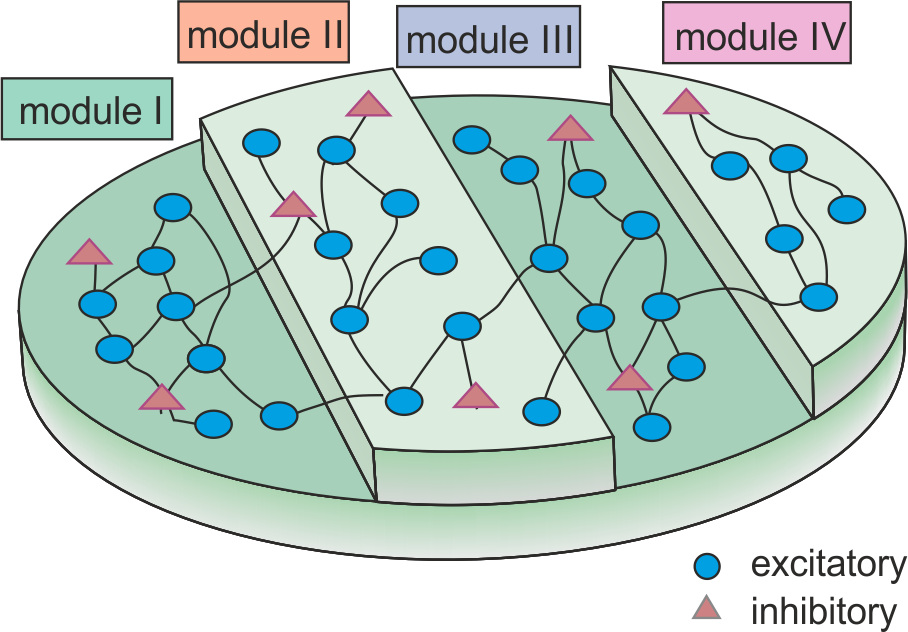} \\
			B \\
			\includegraphics[width=0.2375\linewidth, trim=0 0 0 0, clip]{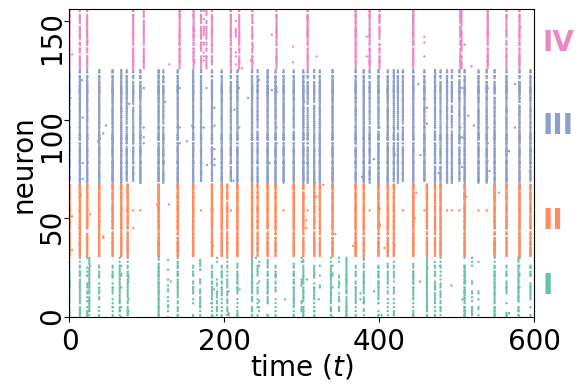} \\
		\end{tabular}
		&
		\hspace{0em}	\begin{tabular}{ll}
			\vspace*{0.2em}C & D\\
			\vspace*{1em}\includegraphics[width=0.35\linewidth, trim=0 0 0 0, clip]{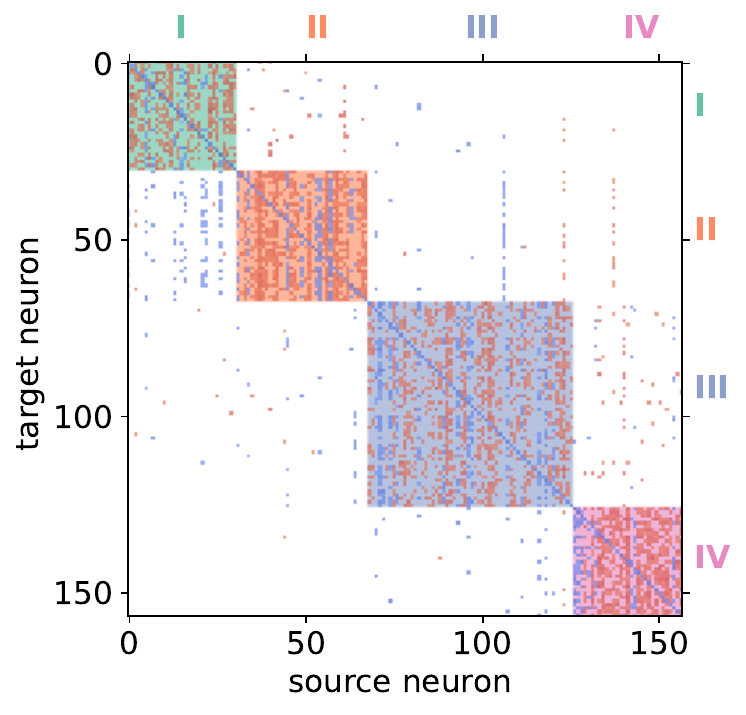} & \raisebox{0.3em}{ \includegraphics[width=0.334\linewidth, trim=0 0 0 0, clip]{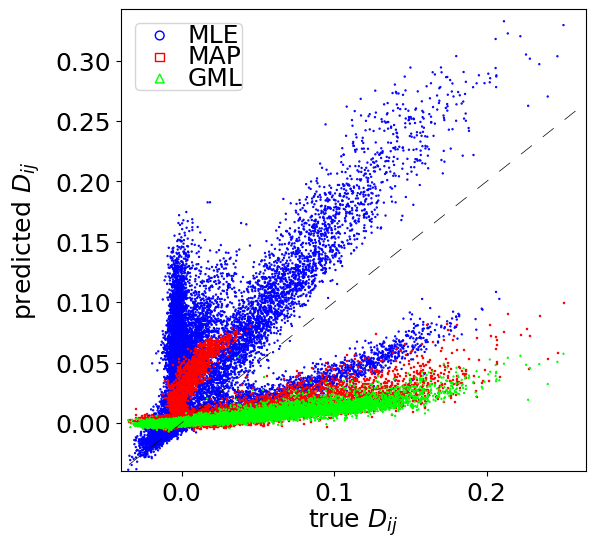}  } \\
		\end{tabular}

	\end{tabular}
	\caption{
		A: A three-dimensional figure illustrating the \emph{in silico} experimental setup. The neurons are lying on stripes (modules I to IV) of patterned substrates, which suppress cross-connections between different stripes. B: The raster plot displays neuronal activities, with each neuron's color corresponding to the module it belongs to. C: The inferred structure $\boldsymbol{J}$ of the \emph{in silico} model obtained using GML and represented by a connectivity matrix. The background color is a mixture of two colors, with each color corresponding to the module of the source and target neurons. Each entry corresponds to the coupling strength $J_{ij}$. A positive (negative) strength colored in red (blue) refers to excitatory (inhibitory) signals sent from source ($i$) to target ($j$) neurons. D: The predicted delayed time covariance $\boldsymbol{D}$ against the true values evaluated from data.
	}
	\label{fig_silico}
\end{figure*}

Since validating the accuracy of neuronal type classification and link existence is very costly and extremely difficult for \emph{in vitro} experiments, we first test our algorithm on emulated \emph{in silico} data since ground-truth topology is known. We generated an \emph{in silico} neuronal network in which the culture is lying over a striped topographical patterned substrate~\cite{montala2022rich} as illustrated in \fig{fig_silico}A. The patterned substrate facilitates high connection density between neurons located on the same stripe but allow for lower connection density across stripes, which is an interesting feature that can be expressed and tested using our method. Spontaneous neuronal activities are then generated on this network. We note that connectivity within a stripe is so strong that each one effectually shapes a `module' of highly interacting neurons. The detailed methodology of how the data is generated is described in the Method section.

Figure~\ref{fig_silico}B shows the raster plot of spontaneous neuronal activities generated \emph{in silico}, with colors indicating the respective module neurons belong to. Using the activities as input, we infer the effective structure $\boldsymbol{J}$ using Maximum Likelihood Estimator (MLE), MAP and GML. Their methodology is described in detail in the Method section and SI. For the prior distributions imposed in MAP and the evidence approximation, we first fix the hyperparameter $a=0.1$, the fraction of excitatory neurons, $\gamma=0.8$ and $\theta_{ij} = 0.5^{m_d}$, where $m_d$ is the number of stripes that separate neurons $i$ and $j$. The values employed for the other hyperparameters used both for MAP inference and as initial values for the GML inference, are obtained using $\boldsymbol{J}$ and $\boldsymbol{H}$ inferred by MLE. The value of $a$ reflects the fact that neurons are less likely to be connected if they are far away from each other; $\gamma$, the ratio between excitatory and inhibitory neurons is already statistically known; while for $\theta_{ij}$, we exploit available information about the physical structure and the value of $0.5$ is chosen arbitrarily. We tested the performance of the inference methods using other values of $a$, $\gamma$ and $\theta_{ij}$ and the results are similar. This is because when the number of samples is large enough, the effect of the priors is suppressed.

The inferred connectivity matrix of $\boldsymbol{J}$ using GML is shown in~\fig{fig_silico}C, where the background mask is the mixture of colors corresponding to the source and target neurons, while inhibitory and excitatory transmissions are colored in blue and red, respectively. We note that the diagonal entries are mostly inhibitory; this does not necessarily correspond to physical connections but may reflect the fact that neurons are less likely to spike again after firing due to their refractory period. Another key feature is that, except for the diagonal, signals in the same column share the same state, which corresponds to the inferred excitatory/inhibitory nature of the neurons; this is validated by comparing the inference results to the true \emph{in silico} model. 

The most appropriate measures of success when contrasting inferred (`infer') and ground-truth (`true') topologies are the positive predictive value (PPV) and negative predictive value (NPV), or precision of neuronal type $P(z_{\mbox{true}}| z_{\mbox{infer}})$, compared to the prior-based random guess, where excitatory nature is the positive case and inhibitory the negative one, as shown in \tab{tab_neuronal_nature}. While being less relevant due to the biased nature of the variables, the true positive rate (TPR -  sensitivity) and the true negative rate (TNR -  specificity), $P(z_{\mbox{infer}}= +/- | z_{\mbox{true}}= +/-)$ are also presented alongside the random guess, for completeness. Overall predictive performance is summed over both cases with the respective probabilities. Notably, no existing method can infer the excitatory and inhibitory nature of neurons from single spontaneous activity recordings without interference, such as channel blocking. With a significant improvement over a prior-based random guess, our method identifies well the individual neuronal type.

\begin{table}
\begin{center}
	\begin{tabular}{||c| c|c| c ||} 
		\hline
	 \textbf{Measure }& \textbf{ \red{TPR}/\blue{TNR} } & \textbf{\red{PPV}/\blue{NPV}}& \textbf{Random}\\ [0.5ex] 
		\hline\hline
		\textbf{Excitatory} & \red{0.88} & \red{\textbf{0.96}}  & 0.8\\ [0.5ex] 
		\hline
	\textbf{Inhibitory} & \blue{0.84}  & \blue{\textbf{0.64}} & 0.2 \\ 
		\hline
\textbf{Overall}& n/a &   \textbf{0.87} & 0.68  \\[0.5ex] 
		\hline
	\end{tabular}
\end{center}
\caption{Success measures in identifying neuron type, true positive rate (TPR -  sensitivity), True negative rate  (TNR -  specifibility) and positive predictive value (PPV) of the \emph{in silico} model study.}
\label{tab_neuronal_nature}
\end{table}

\begin{figure}
	\centering
	\begin{tabular}{l}
		\large{A} \\
		\hspace{1.5em}\includegraphics[width=0.487288136\linewidth, trim=0 0 0 0, clip]{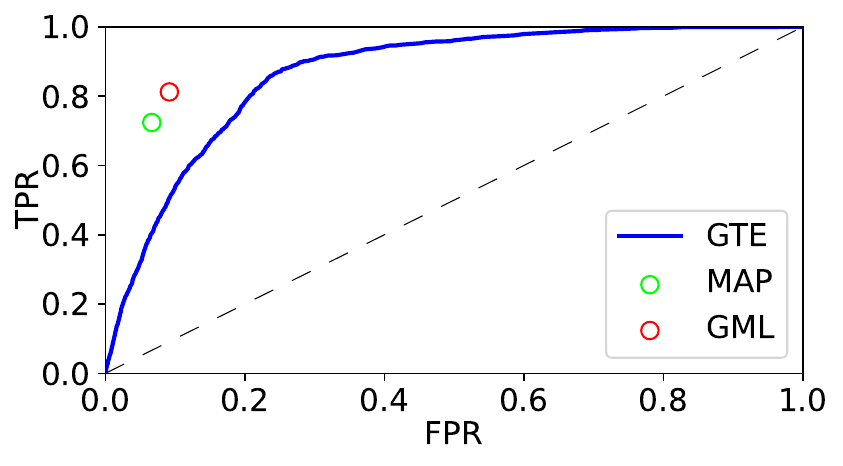} \\
		\large{B} \\
		\hspace{0.1em}\includegraphics[width=0.5\linewidth, trim=0 0 0 0, clip]{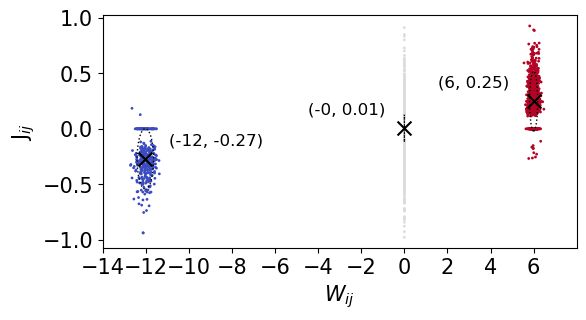} \\
		\large{C}\\
		\hspace{0.95em}\includegraphics[width=0.484550562\linewidth, trim=0 0 0 0, clip]{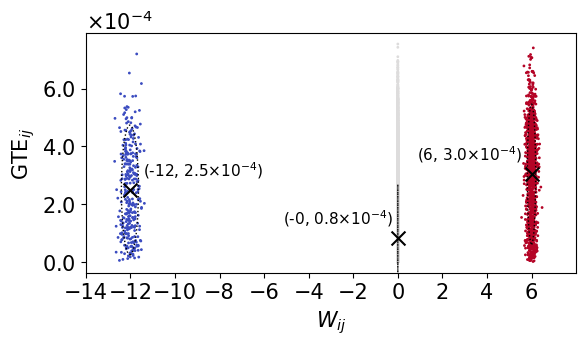}
	\end{tabular}
	\caption{
		A: The receiver operating characteristic (ROC) curve plotting the true positive rate (TPR) against the false positive rate (FPR) of identifying effective links for \emph{in silico} experiments using generalized transfer entropy (GTE continuous blue line). The TPR and TFR using our MAP and GML are marked as green and red nodes respectively. The dashed diagonal line in A refers to a random guess in this case. B and C: The scatter plot of inferred coupling strength $J_{ij}$ and generalized transfer entropy (GTE) against the true synaptic strength of the \emph{in silico} model, respectively. The inhibitory, excitatory and non-existing links are colored in blue, red and gray respectively. 
	}
	\label{fig_silico_W_J_GTE}
\end{figure}

Figure~\ref{fig_silico}C reveals a strong clustering effect for connectivity within modules, indicating that two neurons within the same module have a higher probability of being connected compared to two neurons from different modules. This suggests that the GML captures the structure of the topographical substrate. We then test the performance of inferring  existing links using MAP, GML and GTE~\cite{orlandi2014transfer}. Since GTE is based on an assigned transfer entropy threshold value for identifying links, we plot the complete receiver operating characteristic (ROC) curve for all threshold values, of TPR against FPR of identifying non-zero links as shown in~\fig{fig_silico_W_J_GTE}A. The TPR and FPR of finding non-zero links using GML are $54\%$ and $2.6\%$, respectively, and $57\%$ and $3.3\%$ for MAP, as indicated by the red and green circles, respectively. These values exceed the ROC curve generated by GTE, indicating that our method offers a higher TPR than GTE at the same FPR level, or a lower FPR at the same TPR level. We remark that, although MAP has a higher TPR than that of GML, its FPR is also higher. Additionally, we observe that the difference between MAP and GTE at the same FPR is lower than that of GML, which emphasizes the importance of hyperparameter optimization. It is essential to point out that our approaches detect link existence in a principled manner, eliminating the need for heuristic threshold decisions and rendering the inference results more reliable.

A key feature that our method offers is the inference of the effective synaptic (coupling) strength between neurons, which is crucial for understanding the learning process in both neuroscience and cortical neurons-based learning machines. For example, by observing changes in synaptic strengths over extended periods allows one to study neuronal network plasticity due to stimulation and facilitates the design of efficient stimulation protocols in cortical neurons-based learning devices. We note that the neuronal spiking mechanism of the \emph{in silico} emulator follows the Izhikevich model~\cite{izhikevich2003}, which is distinct from the kinetic Ising model, so that there is no direct mapping between the inferred ($J_{ij}$) and true synaptic strength ($W_{ij}$, actual values used in emulator simulations). To compare the capability of our method with existing techniques, we plot the GML inferred weights $J_{ij}$ and GTE values of each link against the true synaptic strengths $W_{ij}$ in~\fig{fig_silico_W_J_GTE}B and C, respectively. In the true model, the mean value of the inhibitory synaptic strength is $-12$, considerably higher than the excitatory synaptic strength's mean value of $6$. GTE as a specific manifestation of transfer entropy only offers non-negative scores and therefore cannot directly distinguish inhibitory and excitatory links. 

Figure~\ref{fig_silico_W_J_GTE}C, shows the mean value of GTE in the inhibitory group to be $2.6\times 10^{-4}$, which is lower than the value in the excitatory group, $3\times 10^{-4}$ — contradicting the true values. On the other hand, as shown in \fig{fig_silico_W_J_GTE}B, the mean value of the inferred $J_{ij}$ for the inhibitory connections is $-0.27$, which is of higher magnitude than that of excitatory connections, $0.25$. These results suggest that the inferred $J_{ij}$ using our method agrees with the true model synaptic strengths ($W_{ij}$). The GTE values also exhibit a significant variance compared to the magnitudes of the mean values.  Notably, the mean value of the inferred $J_{ij}$ for the missing links is very close to zero with a small variance, indicating that although some of the links are classified as non-zero, incorrectly, the majority are overwhelmingly close to zero. This false positive classification may result from the Gaussian approximation of the delta function in the prior probability $p\left(J_{ij}\right)$. This suggests that the inferred effective coupling strength using our GML method agrees with the true model.

Another advantage of our method is that one can adopt the inferred structure $\boldsymbol{J}$ and $\boldsymbol{H}$ for generating artificial data through Monte Carlo simulations to predict neuronal activity or validate how well the inferred structure describes the true model by comparing quantities of interest, such as the delayed time covariance $\boldsymbol{D}=\left\{ \left\langle s_{i}^{t}s_{j}^{t-1}\right\rangle _{t}\right\} _{i,j}$. Thus, we employed the inferred structures using MLE, MAP and GML to generate artificial neuronal activities and evaluate the predicted delayed time covariance $\boldsymbol{D}$ to its true values, as shown in~\fig{fig_silico}D. We can see that the predicted $\boldsymbol{D}$ values closely align with the true values for all methods, suggesting that the kinetic Ising model effectively explains \emph{in silico} neuronal activities. 

Our GML approach demonstrates strong performance in inferring the neuronal types and link existence as well as the ability for activity prediction. Notably, by comparing the results between MLE and our GML approach, we observe that GML exhibits a gentler slope and deviates more from the perfect prediction manifested by the $y=x$ line. However, our results still demonstrate a strong overall agreement. This difference may due to the fact that MLE does not impose any restrictions on the choice of $J_{ij}$, aiming to provide the best possible description of the data. On the other hand, in GML, the prior probability $p\left(J_{ij}\left|0, z_j\right.\right) \approxeq \mathcal{N}\left(0,\epsilon\right)$ acts as a regularization term and constrain $J_{ij}$ to have the same sign for each fixed $j$. As a result, the inferred $J_{ij}$ values are lower than those obtained through MLE, leading to an underestimation of $\boldsymbol{D}$. Nonetheless, this approach allows for accurate predictability of neuronal types and effective links, which is more important in neuroscience research. 

The above results show that our GML approach performs very well on data generated by the \emph{in silico} model with patterned substrates. In the SI, we study the performance of our GML approach on another \emph{in silico} homogeneous neuronal network with no patterned substrates. We show that even in a homogeneous network, which is harder for structure inference, as there are less constraints for the solution space of $\boldsymbol{J}$, our GML approach still performs very well. The results obtained by the GML for \emph{in silico} data supports the view that it is highly suitable candidate for analyzing biological neuronal network data.

\begin{figure*}
	\centering
	
	\begin{tabular}{l @{} l @{} }
		\begin{tabular}{ll}
			A & B\\
			\raisebox{2.6em}{\includegraphics[width=0.28178\linewidth, trim=0 0 0 0, clip]{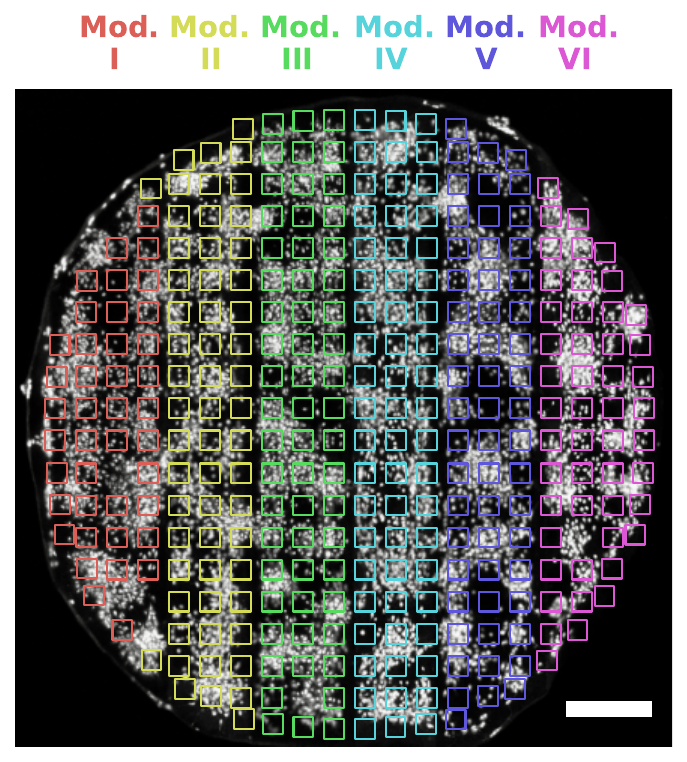} }& \includegraphics[width=0.35\linewidth, trim=0 0 75 0, clip]{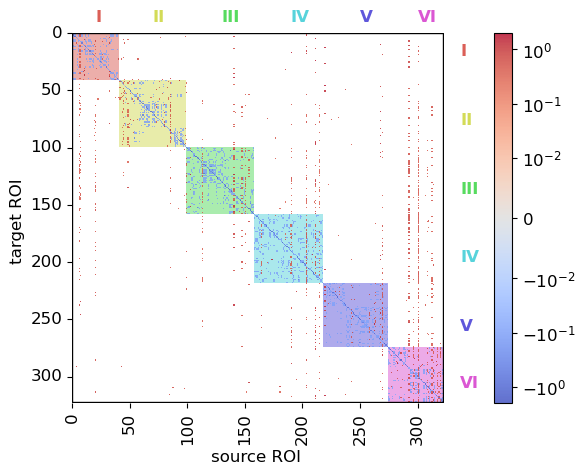} \\
		\end{tabular}
		&
		\hspace*{0em}	\raisebox{-0.3em}{\begin{tabular}{l}
			C \\
			\includegraphics[width=0.28\linewidth, trim=0 9 0 0, clip]{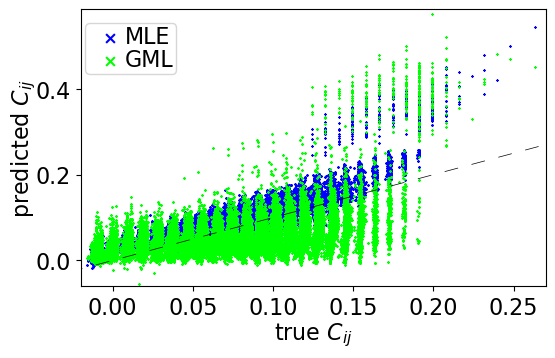} \\
			\raisebox{0em}{D} \\
			\raisebox{0em}{\includegraphics[width=0.28\linewidth, trim=0 9 0 7, clip]{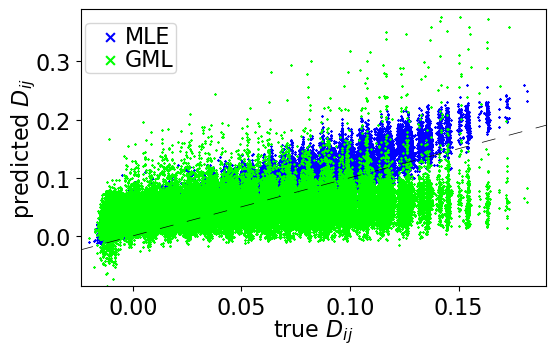} }\\
		\end{tabular}}
	\end{tabular}
	\caption{A: The studied rat cortical neuronal network over striped a topographical pattern $2$~mm in diameter, grouped into different regions of interest (ROIs). ROIs within different stripes are shown as squares with different colors. The scale bar of 200µm is displayed in white in the figure. B: The connectivity matrix of the effective structure inferred from the \emph{in vitro} neuronal activities, using the proposed GML method. Each entry shows the coupling strength of the connections, a negative (positive) strength represents inhibitory (excitatory) connection and is colored in blue (red). The background shade of the plot is the module that the ROI belongs to, with mixed colors where ROIs from two different stripes are connected. A strong clustering effect suggests a dense connectivity of ROIs within stripes and sparse connectivity across stripes. C and D: The predicted equal time covariance $\boldsymbol{C}$ and delayed time covariance $\boldsymbol{D}$ matrices, plotted against the true values evaluated from the data, respectively. }
	\label{fig_invitro}
\end{figure*}

\subsubsection{Experimental data - \emph{in vitro} network with patterned substrates}
Having validated the efficacy of our GML approach to both synthetic data (see SI) and emulator data, we now apply it to an \emph{in vitro} rat cortical neuronal network grown on topographical patterned substrates. Data consisted in neuronal activity recording from calcium fluorescence imaging, processed to consider regions of interest (ROIs) as nodes in the neuronal network, as shown in \fig{fig_invitro}A. Details of data acquisition and analysis are provided in the Methods section.  

Using this experimental data, we infer the effective structure and recreate the neuronal activities for validation. Similar to the \emph{in silico} case, we group ROIs into modules according to the stripes patterning (\fig{fig_invitro}A); the inferred coupling strengths are visualized in the connectivity matrix shown in \fig{fig_invitro}B. The background colors indicate which module the source ROIs belong to. Entries colored in red (blue) indicate effective excitatory (inhibitory) connections from source ($i$) to target ($j$) ROIs. We observe dense connectivity between ROIs from the same stripe and sparse connections across stripes. Furthermore, the likelihood of connection decreases as the distance between ROIs increases. This agrees with the biological understanding that long-range connections are rare to minimize wiring cost, so that neurons preferentially connect to their neighbors, as observed in previous studies using similar patterned substrates~\cite{montala2022rich}. 

While conventional approaches often yield sets of scores, like GTE, they lack a direct measure of estimation quality. Our probabilistic model-based inference, however, allows for the validation of inferred effective structures by reproducing neuronal activities using Monte Carlo simulation. We thus simulated activities using the estimated reconstructed effective network parameters $\boldsymbol{J}$ and $\boldsymbol{H}$ and compared them to the experimental values, evaluating afterwards the equal and delayed time covariance matrices $\boldsymbol{C}$ and $\boldsymbol{D}$, as shown in \fig{fig_invitro}C and D, respectively. Most of the predicted values are aligned with the dashed lines (exact match), meaning a good-quality prediction. However, one can see that for both $\boldsymbol{C}$ and $\boldsymbol{D}$, some of the predicted values are very close to zero. This can be attributed to false positive errors generated due to the restrictions imposed on individual ROIs, such as having the same characteristics, e.g. being either excitatory or inhibitory and sharing the same synaptic connection sign. More realistically, within a single ROI, there might be neurons of different types, leading to some connections being erroneously decimated to zero. We anticipate this problem to be mitigated when smaller ROI sizes are considered and longer calcium recordings are being used, which will be discussed in the next section.

In general, the GML inference method shows excellent agreement between inferred effective network and the underlying  biological structure, with predicted activities fitting well the real data.

\subsubsection{Experimental data - \emph{in vitro} homogeneous network with chemical stimulation}
To complete the results, we considered experiments in which neuronal connectivity was altered through plasticity, which was modulated by using chemical stimulation.

Arguably, the simplest way to understand how neuronal network plasticity takes place due to stimulation, is to compare the effective network structures of a neuronal network before and after stimulation. For that, here we showcase another experiment where we apply our method to an \emph{in vitro} primary homogeneous neuronal network with potassium chloride (KCl) stimulation. Neuronal activity data was also collected through calcium imaging, although here each ROI includes one neuron only. The details of data generation are provided in the Method section.

\begin{figure}
	\centering
	\begin{tabular}{l}
		\large{A} \\
		\hspace{4.3em}\includegraphics[width=0.42\linewidth, trim=0 0 0 0, clip]{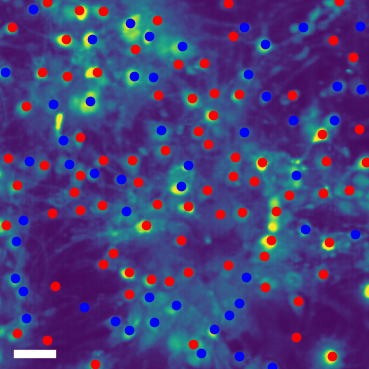} \\
		\large{B} \\
		\hspace{0.3em}\includegraphics[width=0.5\linewidth, trim=0 0 0 0, clip]{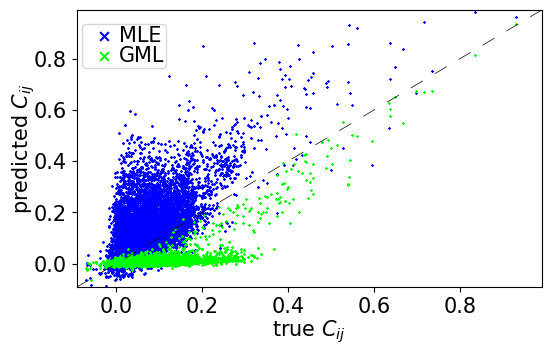} \\
		\large{C}\\
		\hspace{0.3em}\includegraphics[width=0.514\linewidth, trim=0 0 0 0, clip]{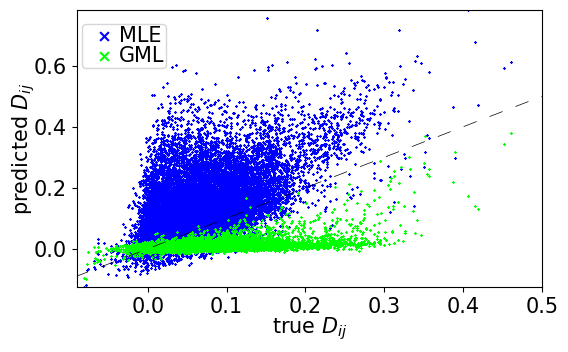}
	\end{tabular}
	\caption{
            A: The \emph{in vitro} primary cortical homogeneous neuronal network analyzed. The red and blue dots correspond to the regions of interest (ROIs) that are classified as excitatory or inhibitory, respectively, where each includes one neuron only. The scale bar of 25µm is displayed in white in the figure.  B and C: The predicted equal and delayed time covariance matrices generated by the inferred structure using Monte-Carlo simulation, $\boldsymbol{C}$ and $\boldsymbol{D}$, respectively, plotted against the true values evaluated from the data.
	}
	\label{fig_vitro_dave}
\end{figure}

We infer the effective structure using the neuronal activity data as input, then recreate the activity measures for validation. We first identify the neuronal type (of single neuron ROIs) as excitatory (red) and inhibitory (blue) as shown in ~\fig{fig_vitro_dave}A. 

In particular, we simulate activities using the inferred $\boldsymbol{J}$ and $\boldsymbol{H}$, then evaluate the equal and delayed time covariance $\boldsymbol{C}$ and $\boldsymbol{D}$ and compare them with the true covariance values, as shown in~\fig{fig_vitro_dave}B and C, respectively. Similar to the patterned substrates case, some predicted values are very close to zero for both $\boldsymbol{C}$ and $\boldsymbol{D}$. However, compared to~\fig{fig_invitro}C and D, we see that the predicted values align better with the dashed line, suggesting a more accurate prediction capability. We note that data shown in~\fig{fig_invitro} and \ref{fig_vitro_dave} are both acquired using calcium imaging approaches, but the resolution of the experiment in~\fig{fig_vitro_dave} enables identification of single neuronal elements. This suggests that by reducing the size of each ROI significantly to consist a single neuron, one can reduce errors with respect to multiple neurons ROIs.

In this section, we have studied two \emph{in vitro} experiments providing different scopes. In the experiment with patterned substrates, we studied a complete extended network; while in the experiment with homogeneous network, we studied neuronal activities under chemical stimulation. While the samples and the experiment conditions are different, it is interesting to see that both covariance matrices show a much higher value under stimulation as the KCl stimulation affects the activity across the sample. A detailed study comparing the neuronal structure and synaptic weights before and after stimulation is underway and is beyond the scope of this work.

\section{Discussion}

Cortical neuronal network inference has long been an open question in neuroscience and is crucial for understanding the underlying mechanisms and properties of neuronal systems. Neuronal cultures are regarded as a promising living model to investigate a broad spectrum of technological challenges, from biologically-inspired AI~\cite{kagan2022vitro,sumi2023biological} to efficient design of treatments for neurological disorders~\cite{Carola2021}. Since the structural blueprint of neuronal connections is not easily accessible in a culture, nor their excitatory/inhibitory nature, indirect techniques to infer such a blueprint have jumped into the front-line of computational neuroscience.

Here we introduced a novel and probabilistic algorithm based on statistical physics and Bayesian techniques for the effective structure inference of biological neuronal networks from firing data. The algorithm can not only infer the effective synaptic strengths between neurons but, more importantly, can identify the excitatory and inhibitory nature of neurons as well as the effective connections between them; this is used in a principled way from single spontaneous recordings without additional interference to the culture such as stimulation. This capability goes beyond what most existing state-of-the-art methods can offer. Through synthetic, \emph{in silico} and realistic \emph{in vitro} experiments, we demonstrate that our algorithm: (1) outperforms existing methods in both synaptic strength inference and effective connections identification; (2) achieves a high accuracy in neuronal type classification; (3) exhibits good reproducibility in the inferred structure, justifying the reliability of the algorithm. In the SI, we also show that our algorithm provides good inference results also in the absence of patterned substrates, in both \emph{in silico} and \emph{in vitro} studies.

The dynamics and functioning of neuronal networks are in large part determined by their connectivity and their evolving synaptic strengths. As such, the method is expected to have a direct impact on neuroscience, cortical neurons-based computing devices, and many other related biological and medical areas. For instance, studying the changes in the effective structure of neuronal cultures over time can lead to new theoretical understandings of how plasticity takes place in response to stimuli. Additionally, gaining insight into information processing and propagation in neuronal networks could greatly impact the development of artificial neuronal networks and neuromorphic computing, e.g., by understanding the importance of the ratio between excitatory and inhibitory neurons on functionality. 

Revealing precise information of effective structure and neuronal types is essential for developing biological machine learning as it helps one to accurately represent the network dynamics, make predictions and test hypotheses. Most importantly, it is critical for the design of stimulation learning protocols. Interdisciplinary research in this direction is underway. 

\section{Method}
\subsection{Preprocessing and optimal time bin determination}
Since the mathematical model we introduce is based on discrete time steps, one needs to binarize the neuronal activities before carrying out the inference process. We note that the determination of the time bin $\tau$ is crucial for inference as the binarized firing patterns can vary significantly with $\tau$. For instance, the delayed time covariance between neurons can vanish if $\tau$ is either too large or small. To address this, we employed an information theory-based method to identify the optimal time bin size $\tau^*$ that optimizes the total mutual information of the system~\cite{terada2018objective, terada2020inferring}. The optimal bin size $\tau^*$ is given by
\begin{align}
	\tau^{*}=\underset{\tau}{\text{argmax}}\left[\left(\frac{T}{\tau}-1\right)\sum_{i\neq j}I_{\tau}\left(\boldsymbol{s}_{i},\boldsymbol{s}_{j}\right)\right],  \label{eq_gross_mutual_info}
\end{align}
where $I_{\tau}\left(\boldsymbol{s}_{i},\boldsymbol{s}_{j}\right)$ is the mutual information between $s_{i}^t$ and $s_{j}^{t-1}$. The idea of mutual information is straightforward: $I_{\tau}\!\!\left(\boldsymbol{s}_{i},\boldsymbol{s}_{j}\!\right)$ measures the discrepancy between the joint probability $P\!\left(s_{i}^{t}\!,s_{j}^{t-1}\!\right)$ and the factorized probability $P\!\left(s_{i}^{t}\right)\!P\!\left(\!s_{j}^{t-1}\right)$ where activities of $i$  and delayed activities of $j$ are assumed to be independent. Thus, a higher $I_{\tau}\!\!\left(\boldsymbol{s}_{i},\boldsymbol{s}_{j}\!\right)$ suggests stronger correlation between neurons $i$ and $j$. Thus, $\tau^*$ maximizes the total mutual information, indicating that the neurons are least likely to be independent of each other and more information about their co-dependence can be extracted. Using $\tau^*$ evaluated in \req{eq_gross_mutual_info} to discretize the neuron firing times into distinct time steps that provide the observed data $\boldsymbol{s}$, which is then ready for the inference process.

\subsection{Inference algorithm}
Unlike the basic kinetic Ising model of statistical physics, inferring the effective structure from neuronal activities faces two major challenges: (1) $\boldsymbol{J}$ is not Gaussian with small variance due to the existence of different neuron types; (2) the connectivity between neurons is affected by multiple factors, including neuronal distance and the patterned substrates (e.g. PDMS stamps). These factors make it difficult to successfully adapt established approaches~\cite{roudi2011mean, mezard2011exact, decelle2015inference, aguilera2021unifying, zeng2011network, zhang2012inference, huang2014dynamics, kappen2000mean} directly to this problem.

\begin{figure}
	\centering
	\includegraphics[width=0.6\linewidth]{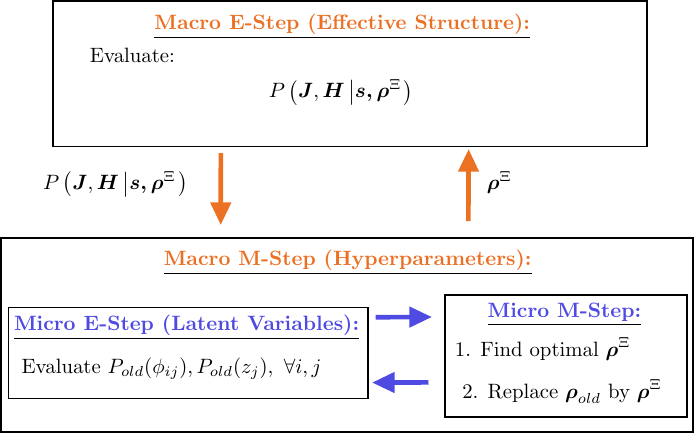}
	\caption{A sketch of the proposed GML-based approximation algorithm. The effective connectivity and the optimal hyperparameters are jointly evaluated by incorporating two applications of the EM algorithm. The posterior distributions of $\boldsymbol{J}$  and $\boldsymbol{H}$ are evaluated in the macro E step, while the hyperparameters values are evaluated in the macro M step, which has an internal nested EM algorithm of its own, representing the interplay between the latent variable values and their distributions. In the micro EM algorithm: the posterior distributions of $\phi_{ij}$ and $z_j$ are evaluated in the micro E step, while the optimal hyperparameters are evaluated in the micro M step, constituting jointly the macro M step that feeds back into the macro E step.}
	\label{fig_emidea}
\end{figure}

To tackle these challenges based on the defined mathematical model we utilize the generalized maximum likelihood (GML)~\cite{bishop2006pattern} technique, also known as evidence approximation, and MAP estimation. This combination enables us to jointly infer optimal hyperparameters, latent variables and the effective network structure in a principled manner. While the detailed derivation is available in the SI, we provide a general overview of the method here.

The aim of this algorithm is to infer the effective structure parameters  $\boldsymbol{J}$  and $\boldsymbol{H}$; this is supported by the optimal $\boldsymbol{\rho}$ that maximize the evidence function $\ln P\left(\boldsymbol{s}\left|\boldsymbol{\rho}\right.\right)$. The optimal values of the latent variables also contribute important insight, determining the type of each neuron (excitatory/inhibitory - $\boldsymbol{z}$) and the existence of links ($\boldsymbol{\phi}$ taking the value $\{0,1\}$ for each link). Our approach involves a two-layered EM algorithm for the estimation of hyperparameters and (latent) variables as illustrated in \fig{fig_emidea} comprising a "macro" and a ``micro" EM algorithms, the latter is being used to perform the M-step of the macro EM algorithm. In the macro E step, the posterior of the effective structure variables, $\boldsymbol{J}$ and $\boldsymbol{H}$, is evaluated. To make the algorithm tractable we focus on the most likely values, determined using gradient descent and the derivatives $\frac{\partial}{\partial J_{ij}}\!\ln P\!\left(\boldsymbol{s}\!\left|\boldsymbol{\rho}\right.\!\right)$ and $\frac{\partial}{\partial H_{i}}\!\ln P\!\left(\boldsymbol{s}\!\left|\boldsymbol{\rho}\right.\!\right)$, given the hyperparamters $\boldsymbol{\rho}$ found in the macro M step. The macro M step maximizes the expected complete-data log likelihood of the evidence function with respect to the hyperparameters $\boldsymbol{\rho}$, to determine the optimal values $\boldsymbol{\rho}^*$ through the secondary EM process; the micro EM process iterates  between the expectation of the probability distributions of the latent variables $\phi_{ij}$ and $z_j$, and maximization of the hyperparameters $\boldsymbol{\rho}$.

Parameters obtained by the macro E step are treated as fixed in the macro M step. In the micro E step, we use the current hyperparameters $\boldsymbol{\rho}_{old}$ (the notation $old$ refers to values obtained from the maximization step in the micro EM iteration part) and the current effective structure $\boldsymbol{J}^*$ and $\boldsymbol{H}^*$ to evaluate the posterior probabilities $P_{old}\left ( \phi_{ij} \right )$ and $P_{old}\left ( z_{j} \right )$ of the latent variables $\phi_{ij}$ and $z_j$. Specifically, the probability distributions of the latent variables are given by
\begin{align}
	P_{old}\left(z_{j}\right)&\propto p\left(\left\{ J_{lj}^{*}\right\} _{l}\left|z_{j},\boldsymbol{\rho}\right.\right)p\left(z_{j}\left|\boldsymbol{\rho}\right.\right), \label{eq_P_old_phi_z} \\
	P_{old}\left(\phi_{ij}\right)&=\sum_{z_{j}}P_{old}\left(\phi_{ij}|z_{j}\right)P_{old}\left(z_{j}\right). \label{eq_P_old_phi} \nonumber
\end{align} 
While in the micro M step, one uses $P_{old}\left ( \phi_{ij} \right )$ and $P_{old}\left ( z_{j} \right )$ evaluated in~Eqs. [\ref{eq_P_old_phi_z},\ref{eq_P_old_phi}] to determine the optimal hyperparameters that maximize the expected complete-data log likelihood
\begin{align}
\mathcal{Q}& =\ln P\left(\boldsymbol{s}\left|\boldsymbol{J},\boldsymbol{H}\right.\right)+\ln p\left(\boldsymbol{H}\left|\vec{\rho}\right.\right)  \\
& +\!\sum_{j}\!\sum_{z_{j}}\!\!\!\sum_{\left\{ \phi_{ij}\right\} _{i}}\!\!p\!\left(\!\left\{ \phi_{ij}\right\} _{i}\!,\!z_{j}\!\left|\!\left\{ \!J_{ij}^{*}\!\right\} \!\!_{i},\boldsymbol{\rho}\right.\!\right)\! \ln p\!\left(\!\!\left\{ J_{ij}^{*}\right\} _{i}\!,\!\left\{ \phi_{ij}\right\} \!_{i}\!,z_{j}\left|\boldsymbol{\rho}\right.\!\right). \nonumber\label{eq_comple_log_likelihood}
\end{align} 
In particular, the hyperparameters $\gamma,\mu_{J}^{+},v_{J}^{+},\mu_{J}^{-},v_{J}^{-},\mu_{H},v_{H}$ are obtained by setting the corresponding derivatives of $\mathcal{Q}$ to zero, while the decay parameter $a$ is estimated using gradient descent.

In general, the inference and optimization  framework consists of iteratively executing the macro E and M steps until convergence. Within each macro M-step, the hyperparameters are evaluated by iteratively conducting the micro E-step and M-step until convergence. Finally, one decides on the structure parameters, neuronal type and link existence by choosing the highest probability states. Notably, for a more sensible starting point, the initial conditions of the hyperparameters $\boldsymbol{\rho}$ can be evaluated using the inferred $\boldsymbol{J}$ and $\boldsymbol{H}$ by using MLE. Additionally, it is worth mentioning that if rapid convergence is preferred over absolute accuracy, the number of macro EM iterations can be limited to one, which effectively results in the MAP estimation.

\subsection{\emph{in silico} data generation}
Emulated neuronal data has been generated with a spiking neuronal network model previously used to model the network growth and activity of biological neuronal culture \cite{alvarez2009,orlandi2013}. 
The existing network growth model has been adapted to incorporate the effect of inhomogeneous environments on the network connectivity thus making the emulated data replicate experimental calcium-recorded results on biological neuronal cultures in inhomogeneous environments \cite{montala2022rich}.

Briefly, network growth is modeled by placing $n$ neurons in a non-overlapping manner on a surface and modeling axon growth from each neuron by concatenating line segments, in which segment $i$ is placed with a random angle $\varphi_i = \varphi_{i-1} + \sigma_{\varphi} \mathcal{N}(0,1)$ with respect to the previous segment $i-1$.
Once an axon segment of neuron $j$ is placed within a radius $r_{soma} \approx 7.5~\text{$\mu$m}$ of another neuron $i$, a connection $W_{ij}$ is made with a probability $\alpha$. 
The strength of the connection is drawn from a Gaussian distribution with a mean and standard deviation depending on the neuron type. 
Inhibitory neurons make up $20\%$ of the network, the remainders are excitatory.

Neuronal dynamics are modeled using the Izhikevich model neuron~\cite{izhikevich2003}, with added synapse dynamics
\begin{equation}
	\begin{aligned}
		\frac{dP_i}{dt} &= -\frac{P_i}{\tau_P} + \beta R_i \delta(v_i - v_{th})  \\
		\frac{dR_i}{dt} &= \frac{1-R_i}{\tau_R} - \gamma R_i \delta(v_i - v_{th}).
	\end{aligned}
\end{equation}
The product $W_{\cdot i} p_i$ represent the post-synaptic potential induced by the neuron $i$, and $R_i$ the corresponding synaptic neurotransmitter reserve.

\subsection{\emph{in vitro} data generation - experimental methods}

\subsubsection{PDMS topographical reliefs}
Topographical patterns were generated by pouring liquid polydimethylsiloxane (PDMS) on specially designed printed circuit board molds shaped as parallel tracks $300$~$\mu$m wide, $70$~$\mu$m high and separated by $200$~$\mu$m~\cite{montala2022rich}. PDMS was cured at 100$^{\circ}$C for 2h, separated from the mold, and perforated with sterile punchers to set 4 PDMS cylinders 2 mm in diameter and 0.5 mm high that contained the inverse topographical pattern of the mold. The 4 cylinders were then evenly distributed on a glass coverslip 13 mm in diameter, autoclaved, and coated with PLL. Flat PDMS substrates were also prepared to investigate the impact of topography.

\subsubsection{Preparation of in-vitro neuronal cultures}
{\em Rat primary cultures for patterned networks on PDMS substrates - }
Sprague-Dawley rat primary neurons (Charles River Laboratories, France) from embryonic cortices at days 18–19 of development were used in all experiments. Manipulation and dissection of the embryonic cortices were carried out under ethical order B-RP-094/15–7125 of the Animal Experimentation Ethics Committee (CEEA) of the University of Barcelona, and in accordance with the regulations of the Generalitat de Catalunya (Spain). Dissection was carried out identically as described in \cite{montala2022rich}. Briefly, cortices were dissected in ice-cold L-15 medium (Gibco), enriched with 0.6\% glucose and 0.5\% gentamicin (Sigma-Aldrich). Brain cortices were first isolated from the meninges and then mechanically dissociated by repeated pipetting. The resulting dissociated neural progenitors were plated on a set of precoated Poly-L-Lysine (PLL, $10$~mg/mL, Sigma-Aldrich) PDMS topographical substrates in the presence of plating medium, which ensured both the development of neurons and glial cells. A density of a half cortex per  $1.3$~cm$^2$ (glass coverslip area) was seeded. This step corresponded to day {\em in vitro} (DIV) 0. Two hours after plating, cells were transduced with adeno-associated viruses bearing the genetically encoded calcium fluorescence indicator GCaMP6s under the Synapsin-I promoter, so that only mature neurons expressed the indicator. At DIV~5 the proliferation of glial cells was restricted by incorporating 0.5\% FUDR in the culture medium for two more days. From DIV~7 onwards, cells were maintained in minimum essential medium supplemented with horse serum (Sigma). This medium was changed periodically every 3 days. Cultures were incubated at 37$^{\circ}$C, 5\% CO$_2$, and 95\% humidity.

{\em Mouse primary neuronal cultures  for network inference - }
13~mm glass coverslips (VWR) were sterilised in 70\% Ethanol for 30 minutes. Coverslips were then transferred to a biological safety cabinet dried completely before coating for 2 hours with 0.02\% Poly-L-Ornithine (Sigma). This was washed once with sterile ddH2O, and then 20~$\mu$g/mL murine laminin was added overnight. P0-P2 C57BL/6 mice were used for were used for the network inference study. All animal procedures were approved by local ethical review and performed in accordance with the United Kingdom Animals Scientific procedures act of 1986 and current EU legislation. The 3 Rs, replacement, and refinement and reduction were considered for planning all animal procedures. The experiment was carried out at Aston University. Cortices were dissected in ice cold HBSS (Gibco) containing 1\% Penicillin/streptomycin (P/S). Meninges were removed before dissection of both cortices, which were then each cut into 8 pieces for enzymatic digestion. Cortices were transferred into 37$^{\circ}$C prewarmed HBSS containing 25~U/mL papain (Sigma), 2~$\mu$g/mL DNAse (Sigma) and L-Cysteine (Sigma). These were incubated at 37$^{\circ}$C for 30 minutes, gently moving rocking the tube every 7.5 minutes. Papain solution was removed, and washed twice for 5 minutes each, at 37$^{\circ}$CC  in MEM (Gibco) + 10\% Horse serum (Sigma) + Glutamax™ (Gibco) + 1\% P/S. Cortices were then mechanically dissociated with glass, fire polished pipettes, to produce homogenous cell suspension. Cells were passed through a 70µm cell strainer (Appleton Woods) to remove any remaining meninges or large clusters of cells or debris. Cells were counted, and seeded onto Poly-L-Ornithine/MuLAM coated glass coverslips at a density of 400,000 cells/cm$^2$. After 2 hours media was replaced with Neurobasal (Gibco) + 1\% B27 (Gibco) + 1\% Glutamax™ (Gibco) + 1\% P/S. Cells were fed every 4 days with a half media change.

\subsubsection{Intracellular calcium fluorescence imaging}
Rat neuronal cultures shaped as $\emptyset$ 2mm PDMS discs allowed the monitoring of the whole network along development. Spontaneous neuronal network activity was recorded using wide-field fluorescence microscopy in combination with the GCaMP6s indicator. Although the networks contained both neurons and glia, only neurons were visualized. Recordings were carried out at DIV~17 for $15$~min on a Zeiss Axiovert C25 inverted microscope equipped with a high–speed camera (Hamamatsu Orca Flash 4.0) in combination with an optical zoom. Recordings were carried out at room temperature with the camera software Hokawo 2.10 at 33 frames per second (fps), 8-bit greyscale format, and a size of $1,024 \times 1,024$ pixels.

For mouse neuronal cultures, cells were loaded with 10~$\mu$M Fluo4-AM in DMSO (Invitrogen) for 40 minutes at 37$^{\circ}$C. Coverslips were then transferred onto an upright Nikon FN1 microscope and images were acquired using a Crest Optics XLight V3 spinning disk confocal and a Teledyne Photometrics Kinetix high speed camera, and in an area of $300 \times 300$~$\mu$m$^2$. The set up was controlled through Micro-Manager~\cite{Edelstein10Manager}.  Cultures were perfused with 37$^{\circ}$C heated Artificial CSF (aCSF) as a control, and an increase of 2.5~mM KCl to increase baseline activity. Cultures were settled for 5 minutes before recording commenced at 10~Hz for 10 minutes. aCSF solution containing the following (in mM): NaCl 120, NaHCO3 25, KCl 2, KH2PO4 1.25, MgSO4 1, and CaCl2 2. aCSF chemicals were obtained from Sigma-Aldrich~\cite{Butcher2022}.

\subsubsection{Data analysis}
Calcium fluorescence recordings were analyzed with the NETCAL software~\cite{orlandi2017netcal,orlandi2017netcal_github} run in MATLAB in combination with custom-made packages. To analyze the data, and as described in \cite{montala2022rich}, Regions of Interest (ROIs) were first laid on the area covered by each culture. For rat primary cultures, ROIs were shaped as a $20 \times 20$ grid centered at the culture and extending its entire $2$~mm circular shape, while for mouse primary cultures ROIs corresponded to individual neurons. Next, the average fluorescence trace within each ROI was extracted as a function of time, corrected from drifts, and normalized. Sharp peaks in the fluorescence signals revealed neuronal activations, which were detected using the Schmitt trigger method, finally leading to binarized time series of neuronal activity in which `1’ indicated the presence of neuronal activity and `0’ its absence.

\subsubsection{Immunocytochemistry}
This technique was used to identify the position of neuronal cell bodies in culture and extract their fluorescence trace with precision. Neuronal cultures were fixed for 20 min with 4\% PFA (Sigma) at room temperature. After washing with PBS the samples were incubated with blocking solution containing 0.03\% Triton (Sigma) and 5\% normal donkey serum (Jackson Immunoresearch) in PBS for 45 min at room temperature. To visualize the neuronal nuclei, the samples were incubated with primary antibodies, against the neuronal marker NeuN (M1406, Sigma), diluted in blocking solution and incubated overnight at 4$^{\circ}$C. Cy3-conjugated secondary antibody against rabbit (711-165-152, Jackson Immunoresear)  was diluted in blocking solution and incubated for 90 min at room temperature. Then, cultures were rinsed with PBS and mounted using DAPI-fluoromount–G (ShouternBiotech). Immunocytochemical images were acquired on a Zeiss confocal microscope.

\section*{Acknowledgements}
This research is supported by the European Union Horizon 2020 research and innovation program under Grant No. 964977 (project NEU-CHiP). JS also acknowledges financial support from the Spanish Ministerio de Ciencia e Innovación under project PID2022-137713NB-C22 and by the Generalitat de Catalunya under project 2021-SGR-00450. The authors acknowledge the support of Aston University Biomedical Facility for the purpose of providing infrastructure support within the College of Health and Life Sciences.


\renewcommand{\appendixname}{Supplementary Information}

\appendix
\beginsupplement
\renewcommand{\thesection}{\Alph{section}}
\renewcommand{\theequation}{S\arabic{equation}}

\section{Generalized maximum likelihood for effective network inference}
In this section, we detail the derivation of a variant of the {\em{generalized maximum likelihood}} (GML), also known as Type II maximum likelihood or evidence approximation. Consider a neuronal network with $N$ interacting neurons. Denote a discrete variable $s_i^t = \{+1,-1$\} when neuron $i$ spikes or is silent at a discrete time step $t$, respectively, for $i=1,\dots,N$. The conditional probability of the neuronal activities of all neurons at time $t$, given the activities of all neurons at the previous time step $t-1$, is as follows:
\begin{align}
	P\left(\boldsymbol{s}^{t}\left|\boldsymbol{s}^{t-1},\boldsymbol{J},\boldsymbol{H}\right.\right)&=\prod_{i=1}^{N}P\left(s_{i}^{t}\left|\boldsymbol{s}^{t-1},\boldsymbol{J},H_{i}\right.\right) =\prod_{i=1}^{N}\frac{\exp\left[\left(H_{i}+\underset{j}{\sum}J_{ij}s_{j}^{t-1}\right)s_{i}^{t}\right]}{2\cosh\left(H_{i}+\sum_{j}J_{ij}s_{j}^{t-1}\right)} \nonumber\\
	&=\exp\left\{ \sum_{i}^{N}\left[\left(H_{i}+\underset{j}{\sum}J_{ij}s_{j}^{t-1}\right)s_{i}^{t}-\ln\left[2\cosh\left(H_{i}+\sum_{j}J_{ij}s_{j}^{t-1}\right)\right]\right]\right\}  \label{eq_trans_prob}~.
\end{align}
Here, $\boldsymbol{J}=\left\{J_{ij}\right\}$, and each $J_{ij}$ denotes the directed coupling strength from neuron $j$ to neuron $i$. The coupling strength can be understood as the effective synaptic strength from neuron $j$ to neuron $i$. Additionally, we denote $H_i$ as the external local field acting on neuron $i$, where the term $H_i-\sum_jJ_{ij}$ can be interpreted as the baseline activity of neuron $i$ when all neighboring neurons of $i$ are silent in the previous time step. 

\begin{figure*}
	\centering
	\includegraphics[width=0.7\linewidth, trim=0 0 0 0, clip]{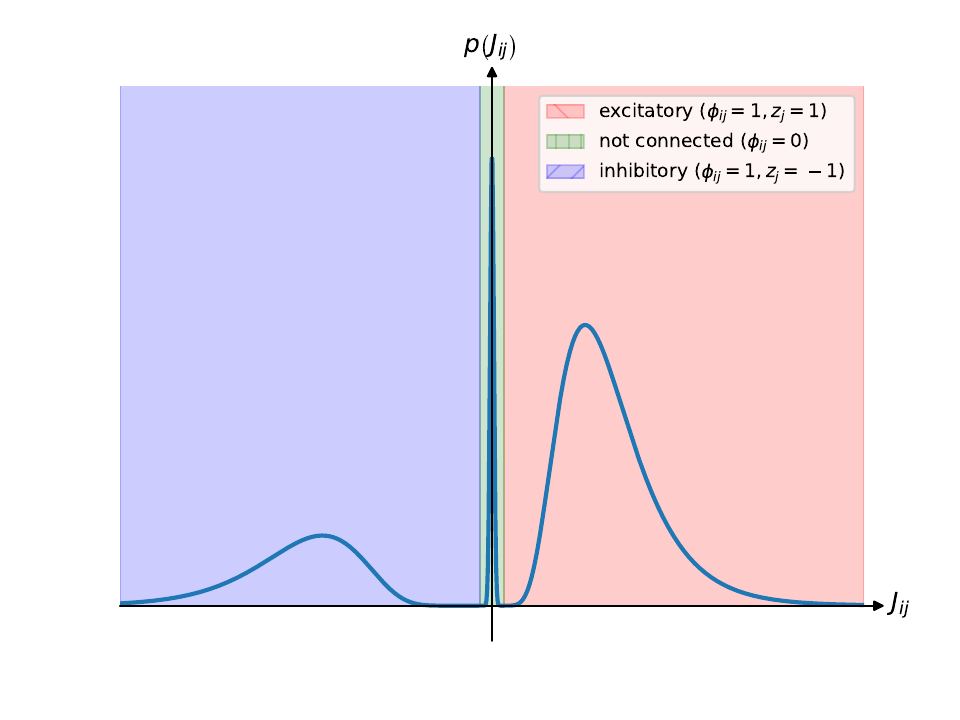} \\
	\caption{
		A sketch illustrating the shape of the prior distribution $p\left(J_{ij}\right)$ for the directed coupling strength. $P\left(J_{ij}\right)$ follows a mixture of three distributions: a log-normal distribution with positive mean to represent the excitatory connections ($\phi_{ij}=1$, $z_j=1$), as shown in the red shaded region; a log-normal distribution with negative mean to represent the inhibitory connections ($\phi_{ij}=1$, $z_j=-1$), as shown in the blue shaded region; a Gaussian distribution with zero mean and small variance, approximately a Dirac delta function, to represent absent connections ($\phi_{ij}=0$), as shown in the green shaded region.
	}
	\label{fig_J_distribution}
\end{figure*}

Now, let us consider the mathematical model and latent variables defined as in the main text. The variables are as follows: $z_j= \{+1,-1$\} represents the excitatory/inhibitory nature of neuron $j$, respectively; $\phi_{ij}={1,0}$ indicates whether the a directed connection from neuron $j$ to neuron $i$ exists or not; $\gamma \in [0,1]$ represents the fraction of excitatory neurons; $\theta_{ij}$ is the prior probability for the existence of a link from neuron $j$ to neuron $i$; $a\in \mathbb{R^+}$ is a decay parameter; $l_{ij}=l_{ji}$ represents the physical distance between neurons $i$ and $j$; $\mu_J^\pm$ and $v_J^\pm \in \mathbb{R}$ are the mean and variance of the distribution of coupling strength $\boldsymbol{J}$ for excitatory and inhibitory neurons, respectively; $\mu_H$ and $v_H \in \mathbb{R}$ are the mean and variance of the distribution for $\boldsymbol{H}$. We generally denote prior probabilities by lower case $p$, and the log of the prior distributions for all variables can be defined as follows:

\begin{align}
	\ln p\left(z_{i}\left|\gamma\right.\right) & =\delta_{z_{i},+1}\ln\gamma+\delta_{z_{i},-1}\ln\left(1-\gamma\right);\\
	\ln p\left(\phi_{ij}\left|a\right.\right) & =\delta_{\phi_{ij},1}\left(\ln\theta_{ij}-al_{ij}\right)+\delta_{\phi_{ij},0}\ln\left(1-\theta_{ij}e^{-al_{ij}}\right);\\
	\ln p\left(H_{i}\left| \mu_{H}, v_{H} \right.\right) & =\frac{-\left(H_{i}-\mu_{H}\right)^{2}}{2v_{H}}-\ln\sqrt{2\pi v_{H}};\\
	\ln p\left(J_{ij}\left|\phi_{ij},z_{j}, \mu_{J}^{\pm}, v_J^\pm \right.\right) & =\delta_{\phi_{ij},0}\ln\delta\left(J_{ij}\right)+\delta_{\phi_{ij},1}\sum_{\varsigma=\pm1}\delta_{z_{j},\varsigma}\mathbbm{1}_{\varsigma J_{ij}>0}\left[\frac{-\left(\ln\zeta J_{ij}-\mu_{J}^{\zeta}\right)^{2}}{2v_{J}^{\zeta}}-\ln\sqrt{2\pi v_{J}^{\zeta}}\right]\\
	& \approx\delta_{\phi_{ij},0}\left[\frac{-J_{ij}^{2}}{2\epsilon}-\ln\sqrt{2\pi\epsilon}\right]+\delta_{\phi_{ij},1}\sum_{\varsigma=\pm1}\delta_{z_{j},\varsigma}\mathbbm{1}_{\varsigma J_{ij}>0}\left[\frac{-\left(\ln\zeta J_{ij}-\mu_{J}^{\zeta}\right)^{2}}{2v_{J}^{\zeta}}-\ln\sqrt{2\pi v_{J}^{\zeta}}\right] \label{eq_prior_J_j}~.
\end{align}
The coefficient $\theta_{ij}$ describes how likely $i$ and $j$ are connected due to physical constraints (e.g. existence of patterned substrates), which is set to be $1$ for the homogeneous environment; while $\epsilon$ is a small real number to approximate $\delta\left(J_{ij}\right)$ by a standard normal distribution with small variance, such that existing connections admit the log-normal distributions and absent connections are close to zero. As illustrated in \fig{fig_J_distribution}, the prior distribution $p\left(J_{ij}\right)$ follows a mixture of three distributions: two log normal distributions with positive and negative means to represent the excitatory and inhibitory connections, respectively; and a standard normal distribution with small variance to represent the absent connections. Log normal distributions are selected for effective connections to fit the biological understanding and act as a constraint to restrict $J_{ij}$ can only be positive (negative) when $z_j = 1$ ($z_j=-1$). For simplicity, we denote $\boldsymbol{\rho}=\left\{ \gamma,a,\mu_{J}^{+},v_{J}^{+},\mu_{J}^{-},v_{J}^{-},\mu_{H},v_{H}\right\} $ as the set of all introduced hyperparameters, where the $\pm$ signs refers to the groups representing excitatory and inhibitory neurons. We remark that $a$ and $\{\theta_{ij}\}$ will be fixed for the whole process with values that agree with biological understanding, which we set to $a=0.1$ in all of our cases. We will optimize the hyperparamters except $a$ and $\{\theta_{ij}\}$ using GML as discussed below.

First, the evidence function, or the marginalized log-likelihood, is given by
\begin{align}
 \ln P\left(\boldsymbol{s}\left|\boldsymbol{\rho}\right.\right)=\ln\int d\boldsymbol{J}~d\boldsymbol{H}~P\left(\boldsymbol{s},\boldsymbol{J},\boldsymbol{H}\left|\boldsymbol{\rho}\right.\right).
\end{align}
Our objective is to maximize the log-likelihood function but do it in stages; we first consider the evidence function of data given the hyperparameters. To achieve this, we apply an Expectation Maximization (EM)-inspired algorithm and call this the macro EM. The expected complete log-likelihood with respect to $\boldsymbol{\rho}$ is defined as 
\begin{align}
\int d\boldsymbol{J}~d\boldsymbol{H}~P\left(\boldsymbol{J},\boldsymbol{H}\left|\boldsymbol{s},\boldsymbol{\rho}^{\Xi}\right.\right)~\ln\left[P\left(\boldsymbol{s},\boldsymbol{J},\boldsymbol{H}\left|\boldsymbol{\rho}\right.\right)\right]
\label{eq_full_exp_comp_L}
\end{align}
In general, we aim to find the maximum log-evidence probability and the posterior probability $P\left(\boldsymbol{J},\boldsymbol{H}\left|\boldsymbol{s},\boldsymbol{\rho}^{\Xi}\right.\right)$ in the macro E-step (originally the expectation probability) by fixing $\boldsymbol{\rho}^{\Xi}$ obtained in the previous macro M-step; and we then evaluate and update $\boldsymbol{\rho}^{\Xi}$ as well as the posterior probabilities of the latent variables, $\boldsymbol{\phi}$ and $\boldsymbol{z}$, in the macro M-step.

\subsection{Macro E-step}
In the macro E-step, we evaluate the posterior probability $P\left(\boldsymbol{J},\boldsymbol{H}\left|\boldsymbol{s},\boldsymbol{\rho}^{\Xi}\right.\right)$, given the the fixed $\boldsymbol{\rho}^{\Xi}$ obtained in the previous macro M-step. Since it is computationally challenging to obtain the full distribution of $P\left(\boldsymbol{J},\boldsymbol{H}\left|\boldsymbol{s},\boldsymbol{\rho}^{\Xi}\right.\right)$, we approximate using the saddle point method by its most likely values,
\begin{align}
	\boldsymbol{J}^{*},\boldsymbol{H}^{*} & =\underset{\left\{ \boldsymbol{J},\boldsymbol{H}\right\} }{\text{argmax}}P\left(\boldsymbol{J},\boldsymbol{H}\left|\boldsymbol{s},\boldsymbol{\rho}^{\Xi}\right.\right) = \underset{\left\{ \boldsymbol{J},\boldsymbol{H}\right\} }{\text{argmax}}\ln P\left(\boldsymbol{J},\boldsymbol{H}\left|\boldsymbol{s},\boldsymbol{\rho}^{\Xi}\right.\right) ,
\end{align}
relying on the large system size, which gives the maximum posterior probability.

Then, we approximate $P\left(\boldsymbol{J},\boldsymbol{H}\left|\boldsymbol{s},\boldsymbol{\rho}\right.\right) \approx\delta\left(\boldsymbol{J}-\boldsymbol{J}^{*}\right)\delta\left(\boldsymbol{H}-\boldsymbol{H}^{*}\right)$ after $\boldsymbol{J}^{*}$ and $\boldsymbol{H}^{*}$ are found. Finding the maximum of $P\left(\boldsymbol{J},\boldsymbol{H}\left|\boldsymbol{s},\boldsymbol{\rho}\right.\right)$ is equivalent to finding the maximum for $\ln P\left(\boldsymbol{J},\boldsymbol{H}\left|\boldsymbol{s},\boldsymbol{\rho}\right.\right)= \ln\sum_{\boldsymbol{\phi}}\sum_{\boldsymbol{z}}P\left(\boldsymbol{J},\boldsymbol{H},\boldsymbol{\phi},\boldsymbol{z}\left|\boldsymbol{s},\boldsymbol{\rho}^{\Xi}\right.\right)$ which is intractable. Instead, we find the optimal $\boldsymbol{J}^{*}$ and $\boldsymbol{H}^{*}$ that maximize the expected complete log likelihood (in the notation, we omit the dependence on $\boldsymbol{s}$ and $\boldsymbol{\rho}^{\Xi}$): 
\begin{align}
	\label{eq_exp_log_like_latent}
	\mathcal{Q}_{\Xi}\left(\boldsymbol{J},\boldsymbol{H}\right) = \sum_{\boldsymbol{\phi}}\sum_{\boldsymbol{z}} P_{\Xi}\left(\boldsymbol{\phi},\boldsymbol{z}\right)\ln P\left(\boldsymbol{J},\boldsymbol{H},\boldsymbol{\phi},\boldsymbol{z}\left|\boldsymbol{s},\boldsymbol{\rho}^{\Xi}\right.\right),
\end{align}
where $P_{\Xi}\left(\cdot\right)$ denotes the posterior probability $P\left(\cdot\left|\boldsymbol{s},\boldsymbol{J},\boldsymbol{H},\boldsymbol{\rho}^{\Xi}\right.\right)$ evaluated in the previous macro M-step, for brevity. Then, simplifying \req{eq_exp_log_like_latent}, we get
\begin{alignb}
    \mathcal{Q}_{\Xi}\left(\boldsymbol{J},\boldsymbol{H}\right)&=\sum_{\boldsymbol{\phi}}\sum_{\boldsymbol{z}}P_{\Xi}\left(\boldsymbol{\phi},\boldsymbol{z}\right)\ln\left[P\left(\boldsymbol{s}\left|\boldsymbol{J},\boldsymbol{H},\boldsymbol{\phi},\boldsymbol{z},\boldsymbol{\rho}^{\Xi}\right.\right)p\left(\boldsymbol{J},\boldsymbol{H},\boldsymbol{\phi},\boldsymbol{z}\left|\boldsymbol{\rho}^{\Xi}\right.\right)/P\left(\boldsymbol{s}\left|\boldsymbol{\phi},\boldsymbol{z},\boldsymbol{\rho}^{\Xi}\right.\right)\right]\\
    &=\sum_{\boldsymbol{\phi}}\sum_{\boldsymbol{z}}P_{\Xi}\left(\boldsymbol{\phi},\boldsymbol{z}\right)\ln\left[P\left(\boldsymbol{s}\left|\boldsymbol{J},\boldsymbol{H}\right.\right)p\left(\boldsymbol{J},\boldsymbol{H}\left|\boldsymbol{\phi},\boldsymbol{z},\boldsymbol{\rho}^{\Xi}\right.\right)p\left(\boldsymbol{\phi},\boldsymbol{z}\left|\boldsymbol{\rho}^{\Xi}\right.\right)/P\left(\boldsymbol{s}\left|\boldsymbol{\phi},\boldsymbol{z},\boldsymbol{\rho}^{\Xi}\right.\right)\right]\\
    &=\ln P\left(\boldsymbol{s}\left|\boldsymbol{J},\boldsymbol{H}\right.\right)+\sum_{\boldsymbol{\phi}}\sum_{\boldsymbol{z}}P_{\Xi}\left(\boldsymbol{\phi},\boldsymbol{z}\right)\ln\prod_{ij}p\left(J_{ij}\left|\phi_{ij},z_{j},\boldsymbol{\rho}^{\Xi}\right.\right)\\
    &\qquad+\sum_{\boldsymbol{\phi}}\sum_{\boldsymbol{z}}P_{\Xi}\left(\boldsymbol{\phi},\boldsymbol{z}\right)\ln\prod_{i}p\left(H_{i}\left|\boldsymbol{\rho}^{\Xi}\right.\right)+\sum_{\boldsymbol{\phi}}\sum_{\boldsymbol{z}}P_{\Xi}\left(\boldsymbol{\phi},\boldsymbol{z}\right)\ln\prod_{ij}p\left(\phi_{ij}\left|\boldsymbol{\rho}^{\Xi}\right.\right)\\
    &\qquad+\sum_{\boldsymbol{\phi}}\sum_{\boldsymbol{z}}P_{\Xi}\left(\boldsymbol{\phi},\boldsymbol{z}\right)\ln\prod_{i}p\left(z_{i}\left|\boldsymbol{\rho}^{\Xi}\right.\right)-\sum_{\boldsymbol{\phi}}\sum_{\boldsymbol{z}}P_{\Xi}\left(\boldsymbol{\phi},\boldsymbol{z}\right)\ln P\left(\boldsymbol{s}\left|\boldsymbol{\phi},\boldsymbol{z},\boldsymbol{\rho}^{\Xi}\right.\right)\\
    &=\ln P\left(\boldsymbol{s}\left|\boldsymbol{J},\boldsymbol{H}\right.\right)+\sum_{ij}\sum_{z_{j}=\pm1}\sum_{\phi_{ij}=1,0}P_{\Xi}\left(\phi_{ij}\left|z_{j}\right.\right)P_{\Xi}\left(z_{j}\right)\ln p\left(J_{ij}\left|\phi_{ij},z_{j},\boldsymbol{\rho}^{\Xi}\right.\right)\\
    &\qquad+\sum_{ij}\sum_{z_{j}=\pm1}\sum_{\phi_{ij}=1,0}P_{\Xi}\left(\phi_{ij}\left|z_{j}\right.\right)P_{\Xi}\left(z_{j}\right)\ln p\left(\phi_{ij}\left|\boldsymbol{\rho}^{\Xi}\right.\right)\\
    &\qquad+\sum_{j}\sum_{z_{j}=\pm1}P_{\Xi}\left(z_{j}\right)\ln\prod_{i}p\left(z_{i}\left|\boldsymbol{\rho}^{\Xi}\right.\right)+\sum_{i}\ln p\left(H_{i}\left|\boldsymbol{\rho}^{\Xi}\right.\right)-\sum_{\boldsymbol{\phi}}\sum_{\boldsymbol{z}}P_{\Xi}\left(\boldsymbol{\phi},\boldsymbol{z}\right)\ln P\left(\boldsymbol{s}\left|\boldsymbol{\phi},\boldsymbol{z},\boldsymbol{\rho}^{\Xi}\right.\right)\\
    &=\sum_{t}\sum_{i}\left\{ \left(H_{i}+\sum_{j}J_{ij}s_{j}^{t-1}\right)s_{i}^{t}-\ln\left[2\cosh\left(H_{i}+\sum_{j}J_{ij}s_{j}^{t-1}\right)\right]\right\} \\
    &\qquad+\sum_{i}\sum_{j}\sum_{z_{j}=\pm1}\sum_{\phi_{ij}=1,0}P_{\Xi}\left(\phi_{ij}\left|z_{j}\right.\right)P_{\Xi}\left(z_{j}\right)\Bigg\{\delta_{\phi_{ij},0}\left[-\left(J_{ij}\right)^{2}/\left(2\epsilon\right)-\ln\sqrt{2\pi\epsilon}\right]\\
    &\qquad\qquad+\delta_{\phi_{ij},1}\sum_{\varsigma=\pm1}\delta_{z_{j},\varsigma}\mathbbm{1}_{\varsigma J_{ij}>0}\left[-\left(\ln\zeta J_{ij}-\mu_{J}^{\zeta,\Xi}\right)^{2}/\left(2v_{J}^{\zeta}\right)-\ln\sqrt{2\pi v_{J}^{\zeta}}\right]\Bigg\}\\
    &\qquad+\sum_{i}\sum_{j}\sum_{z_{j}=\pm1}\sum_{\phi_{ij}=1,0}P_{\Xi}\left(\phi_{ij}\left|z_{j}\right.\right)P_{\Xi}\left(z_{j}\right)\left[\delta_{\phi_{ij},1}\left(\ln\theta_{ij}-al_{ij}\right)+\delta_{\phi_{ij},0}\ln\left(1-\theta_{ij}e^{-al_{ij}}\right)\right]\\
    &\qquad+\sum_{i}\sum_{z_{i}=\pm1}P_{\Xi}\left(z_{j}\right)\left[\delta_{z_{i},+1}\ln\gamma+\delta_{z_{i},-1}\ln\left(1-\gamma\right)\right]\\
    &\qquad+\sum_{i}\left\{ \frac{-\left(H_{i}-\mu_{H}\right)^{2}}{2v_{H}^{\Xi}}-\ln\sqrt{2\pi v_{H}^{\Xi}}\right\} -\sum_{\boldsymbol{\phi}}\sum_{\boldsymbol{z}}P_{\Xi}\left(\boldsymbol{\phi},\boldsymbol{z}\right)\ln P\left(\boldsymbol{s}\left|\boldsymbol{\phi},\boldsymbol{z},\boldsymbol{\rho}^{\Xi}\right.\right).
\end{alignb}
The optimization process is carried out through gradient descent, in particular, 
\begin{align}
	\left(\delta\boldsymbol{J},\delta\boldsymbol{H}\right)=\eta\nabla\mathcal{Q}_{\Xi}\left(\boldsymbol{J},\boldsymbol{H}\right),
\end{align}
where $\nabla\mathcal{Q}_{\Xi}\left(\boldsymbol{J},\boldsymbol{H}\right)=\left(\left\{ \frac{\partial\mathcal{Q}_{\Xi}}{\partial J_{ij}}\right\} _{ij},\left\{ \frac{\partial\mathcal{Q}_{\Xi}}{\partial H_{i}}\right\} _{i}\right)$
and $\eta$ is the decaying learning rate; the gradients are given as 
\begin{align}
	\frac{\partial\mathcal{Q}_{\Xi}}{\partial J_{ij}} & =T\left[\left\langle s_{i}^{t}s_{j}^{t-1}\right\rangle _{t}-\left\langle s_{j}^{t-1}\tanh\left(H_{i}+\sum_{k}J_{ik}s_{k}^{t-1}\right)\right\rangle _{t}\right] \nonumber \\
	& -\sum_{z_{j}=\pm1}\sum_{\phi_{ij}=1,0}P_{\Xi}\left(\phi_{ij}\left|z_{j}\right.\right)P_{\Xi}\left(z_{j}\right)\left\{ \delta_{\phi_{ij},0}\frac{J_{ij}}{\epsilon}+\delta_{\phi_{ij},1}\sum_{\varsigma=\pm1}\delta_{z_{j},\varsigma}\mathbbm{1}_{\varsigma J_{ij}>0}\frac{\ln\zeta J_{ij}-\mu_{J}^{\zeta,\Xi}}{J_{ij}v_{J}^{\zeta}}\right\} ,
\end{align}
and 
\begin{align}
	\frac{\partial\mathcal{Q}_{\Xi}}{\partial H_{i}} & =T\left[\left\langle s_{i}^{t}\right\rangle _{t}-\left\langle \tanh\left(H_{i}+\sum_{j}J_{ij}s_{j}^{t-1}\right)\right\rangle _{t}\right]-\frac{H_{i}-\mu_{H}}{v_{H}^{\Xi}}.
\end{align}
Upon convergence, the optimal $\boldsymbol{J}^{*}$ and $\boldsymbol{H}^{*}$ is fixed for the evaluation of $\boldsymbol{\rho}$ and the posterior probabilities in the following macro M-step.

\subsection{Macro M-step}
Fixing the peak values $\boldsymbol{J}^{*}$ and $\boldsymbol{H}^{*}$ obtained in the previous macro E-step, the log-likelihood $\boldsymbol{\rho}$ becomes
\begin{align}
	\label{eq_exp_log_like_latent_map}
	\int d\boldsymbol{J}~d\boldsymbol{H}~P\left(\boldsymbol{J},\boldsymbol{H}\left|\boldsymbol{s},\boldsymbol{\rho}^{\Xi}\right.\right)\ln\left[P\left(\boldsymbol{s},\boldsymbol{J},\boldsymbol{H}\left|\boldsymbol{\rho}\right.\right)\right] & = \int d\boldsymbol{J}~d\boldsymbol{H}~\delta\left(\boldsymbol{J}-\boldsymbol{J}^{*}\right)\delta\left(\boldsymbol{H}-\boldsymbol{H}^{*}\right)\ln\left[P\left(\boldsymbol{s},\boldsymbol{J},\boldsymbol{H}\left|\boldsymbol{\rho}\right.\right)\right] \nonumber \\
	& = \ln\left[P\left(\boldsymbol{s},\boldsymbol{J}^{*},\boldsymbol{H}^{*}\left|\boldsymbol{\rho}\right.\right)\right] \nonumber\\
	& =\ln \left[ \sum_{\boldsymbol{\phi}}\sum_{\boldsymbol{z}}P\left(\boldsymbol{s},\boldsymbol{J}^{*},\boldsymbol{H}^{*},\boldsymbol{\phi},\boldsymbol{z}\left|\boldsymbol{\rho}\right.\right) \right].
\end{align}
The objective of the macro M-step is to find the optimal value $\boldsymbol{\rho}^{*}$ that maximizes \req{eq_exp_log_like_latent_map}. Interpreting $\boldsymbol{\phi}$ and $\boldsymbol{z}$ as latent variables, we apply a second EM algorithm, termed the micro EM, to obtain $\boldsymbol{\rho}^{*}$. The expected complete data log-likelihood is then given by 
\begin{align}
	\mathcal{Q} & =\sum_{\boldsymbol{\phi}}\sum_{\boldsymbol{z}}P\left(\boldsymbol{\phi},\boldsymbol{z}\left|\boldsymbol{s},\boldsymbol{J}^{*},\boldsymbol{H}^{*},\boldsymbol{\rho}^{\text{old}}\right.\right)\ln P\left(\boldsymbol{s},\boldsymbol{J}^{*},\boldsymbol{H}^{*},\boldsymbol{\phi},\boldsymbol{z}\left|\boldsymbol{\rho}\right.\right) \nonumber \\
	& =\sum_{\boldsymbol{\phi}}\sum_{\boldsymbol{z}}P\left(\boldsymbol{\phi},\boldsymbol{z}\left|\boldsymbol{s},\boldsymbol{J}^{*},\boldsymbol{H}^{*},\boldsymbol{\rho}^{\text{old}}\right.\right)\ln\left[P\left(\boldsymbol{s}\left|\boldsymbol{J}^{*},\boldsymbol{H}^{*},\boldsymbol{\phi},\boldsymbol{z},\boldsymbol{\rho}\right.\right)p\left(\boldsymbol{J}^{*},\boldsymbol{H}^{*},\boldsymbol{\phi},\boldsymbol{z}\left|\boldsymbol{\rho}\right.\right)\right]\nonumber \\
	& =\sum_{\boldsymbol{\phi}}\sum_{\boldsymbol{z}}P\left(\boldsymbol{\phi},\boldsymbol{z}\left|\boldsymbol{s},\boldsymbol{J}^{*},\boldsymbol{H}^{*},\boldsymbol{\rho}^{\text{old}}\right.\right)\ln\left[P\left(\boldsymbol{s}\left|\boldsymbol{J}^{*},\boldsymbol{H}^{*}\right.\right)p\left(\boldsymbol{H}^{*}\left|\boldsymbol{\rho}\right.\right)p\left(\boldsymbol{J}^{*},\boldsymbol{\phi},\boldsymbol{z}\left|\boldsymbol{\rho}\right.\right)\right]\nonumber \\
	& =\ln P\left(\boldsymbol{s}\left|\boldsymbol{J}^{*},\boldsymbol{H}^{*}\right.\right)+\ln p\left(\boldsymbol{H}^{*}\left|\boldsymbol{\rho}\right.\right)+\sum_{\boldsymbol{\phi}}\sum_{\boldsymbol{z}}P\left(\boldsymbol{\phi},\boldsymbol{z}\left|\boldsymbol{s},\boldsymbol{J}^{*},\boldsymbol{H}^{*},\boldsymbol{\rho}^{\text{old}}\right.\right)\ln p\left(\boldsymbol{J}^{*},\boldsymbol{\phi},\boldsymbol{z}\left|\boldsymbol{\rho}\right.\right),
\end{align}
where $\boldsymbol{\rho}^{\text{old}}$ is the values of the hyperparameters evaluated in the previous micro M-step. For simplicity, we denote $P\left(\cdot\left|\boldsymbol{s},\boldsymbol{J}^{*},\boldsymbol{H}^{*},\boldsymbol{\rho}^{\text{old}}\right.\right)$
as $P_{\text{old}}\left(\cdot\right)$. Consider the term 
\begin{align}
\label{eq_exp_log_like_small}
	\sum_{\boldsymbol{\phi}}\sum_{\boldsymbol{z}}P_{\text{old}}\left(\boldsymbol{\phi},\boldsymbol{z}\right)\ln p\left(\boldsymbol{J}^{*},\boldsymbol{\phi},\boldsymbol{z}\left|\boldsymbol{\rho}\right.\right) & =\sum_{\boldsymbol{\phi}}\sum_{\boldsymbol{z}}P_{\text{old}}\left(\boldsymbol{\phi},\boldsymbol{z}\right)\ln\left[p\left(\boldsymbol{J}^{*}\left|\boldsymbol{\phi},\boldsymbol{z},\boldsymbol{\rho}\right.\right)p\left(\boldsymbol{\phi},\boldsymbol{z}\left|\boldsymbol{\rho}\right.\right)\right]\nonumber\\
	& =\sum_{\boldsymbol{\phi}}\sum_{\boldsymbol{z}}P_{\text{old}}\left(\boldsymbol{\phi}\left|\boldsymbol{z}\right.\right)P_{\text{old}}\left(\boldsymbol{z}\right)\ln p\left(\boldsymbol{J}^{*}\left|\boldsymbol{\phi},\boldsymbol{z},\boldsymbol{\rho}\right.\right)\nonumber\\
	& \quad+\sum_{\boldsymbol{\phi}}\sum_{\boldsymbol{z}}P_{\text{old}}\left(\boldsymbol{\phi}\left|\boldsymbol{z}\right.\right)P_{\text{old}}\left(\boldsymbol{z}\right)\ln p\left(\boldsymbol{\phi}\left|\boldsymbol{z},\boldsymbol{\rho}\right.\right) +\sum_{\boldsymbol{z}}P_{\text{old}}\left(\boldsymbol{z}\right)\ln p\left(\boldsymbol{z}\left|\boldsymbol{\rho}\right.\right).
\end{align}
By considering the terms in \req{eq_exp_log_like_small} one by one:
\begin{align*}
	& \sum_{\left\{ \boldsymbol{\phi}\right\} }\sum_{\left\{ \boldsymbol{z}\right\} }P_{\text{old}}\left(\boldsymbol{\phi}\left|\boldsymbol{z}\right.\right)P_{\text{old}}\left(\boldsymbol{z}\right)\ln p\left(\boldsymbol{J}^{*}\left|\boldsymbol{\phi},\boldsymbol{z},\boldsymbol{\rho}\right.\right)= \sum_{\left\{ \boldsymbol{\phi}\right\} }\sum_{\left\{ \boldsymbol{z}\right\} }P_{\text{old}}\left(\boldsymbol{\phi}\left|\boldsymbol{z}\right.\right)P_{\text{old}}\left(\boldsymbol{z}\right)\ln\prod_{j}p\left(\left\{ J_{kj}^{*}\right\} _{k}\left|\left\{ \phi_{kj}\right\} _{k},z_{j},\boldsymbol{\rho}\right.\right)\\
	= & \sum_{j}\left[\sum_{\left\{ \boldsymbol{\phi}\right\} }\sum_{\left\{ \boldsymbol{z}\right\} }P_{\text{old}}\left(\boldsymbol{\phi}\left|\boldsymbol{z}\right.\right)P_{\text{old}}\left(\boldsymbol{z}\right)\ln p\left(\left\{ J_{kj}^{*}\right\} _{k}\left|\left\{ \phi_{kj}\right\} _{k},z_{j},\boldsymbol{\rho}\right.\right)\right]\\
	= & \sum_{j}\sum_{\left\{ \boldsymbol{\phi}\right\} }\sum_{\left\{ \boldsymbol{z}\right\} }\left\{ \prod_{a}\left[P_{\text{old}}\left(\left\{ \phi_{ka}\right\} _{k}\left|z_{a}\right.\right)P_{\text{old}}\left(z_{a}\right)\right]\ln p\left(\left\{ J_{kj}^{*}\right\} _{k}\left|\left\{ \phi_{kj}\right\} _{k},z_{j},\boldsymbol{\rho}\right.\right)\right\} \\
	= & \sum_{j}\sum_{\left\{ \boldsymbol{\phi}\right\} }\sum_{\left\{ \boldsymbol{z}\right\} }\left\{\left[ \prod_{a\neq j} P_{\text{old}}\left(\left\{ \phi_{ka}\right\} _{k}\left|z_{a}\right.\right)P_{\text{old}}\left(z_{a}\right)\right] P_{\text{old}}\left(\left\{ \phi_{kj}\right\} _{k}\left|z_{j}\right.\right)P_{\text{old}}\left(z_{j}\right)\ln p\left(\left\{ J_{kj}^{*}\right\} _{k}\left|\left\{ \phi_{kj}\right\} _{k},z_{j},\boldsymbol{\rho}\right.\right)\right\}\\
	= & \sum_{j}\sum_{\left\{ \phi_{kj}\right\} _{k}}\sum_{z_{j}=\pm1}P_{\text{old}}\left(\left\{ \phi_{kj}\right\} _{k}\left|z_{j}\right.\right)P_{\text{old}}\left(z_{j}\right)\ln p\left(\left\{ J_{kj}^{*}\right\} _{k}\left|\left\{ \phi_{kj}\right\} _{k},z_{j},\boldsymbol{\rho}\right.\right)\\
	= & \sum_{j}\sum_{z_{j}=\pm1}P_{\text{old}}\left(z_{j}\right)\sum_{\left\{ \phi_{kj}\right\} _{k}}P_{\text{old}}\left(\left\{ \phi_{kj}\right\} _{k}\left|z_{j}\right.\right)\ln\prod_{i}p\left(J_{ij}^{*}\left|\phi_{ij},z_{j},\boldsymbol{\rho}\right.\right)\\
	= & \sum_{j}\sum_{i}\sum_{z_{j}=\pm1}P_{\text{old}}\left(z_{j}\right)\sum_{\left\{ \phi_{kj}\right\} _{k}}\prod_{b}P_{\text{old}}\left(\phi_{bj}\left|z_{j}\right.\right)\ln p\left(J_{ij}^{*}\left|\phi_{ij},z_{j},\boldsymbol{\rho}\right.\right)\\
	= & \sum_{j}\sum_{i}\sum_{z_{j}=\pm1}P_{\text{old}}\left(z_{j}\right)\sum_{\left\{ \phi_{kj}\right\} _{k}}\left[\prod_{b\neq i}P_{\text{old}}\left(\phi_{bj}\left|z_{j}\right.\right)\right]P_{\text{old}}\left(\phi_{ij}\left|z_{j}\right.\right)\ln p\left(J_{ij}^{*}\left|\phi_{ij},z_{j},\boldsymbol{\rho}\right.\right)\\
	= & \sum_{j}\sum_{i}\sum_{z_{j}=\pm1}P_{\text{old}}\left(z_{j}\right)\sum_{\phi_{ij}=1,0}P_{\text{old}}\left(\phi_{ij}\left|z_{j}\right.\right)\ln p\left(J_{ij}^{*}\left|\phi_{ij},z_{j},\boldsymbol{\rho}\right.\right)\\
	= & \sum_{i}\sum_{j}\sum_{z_{j}=\pm1}\sum_{\phi_{ij}=1,0}P_{\text{old}}\left(\phi_{ij}\left|z_{j}\right.\right)P_{\text{old}}\left(z_{j}\right)\ln p\left(J_{ij}^{*}\left|\phi_{ij},z_{j},\boldsymbol{\rho}\right.\right)\\
	= & \sum_{i}\sum_{j}\sum_{z_{j}=\pm1}\sum_{\phi_{ij}=1,0}P_{\text{old}}\left(\phi_{ij}\left|z_{j}\right.\right)P_{\text{old}}\left(z_{j}\right)\Bigg\{\delta_{\phi_{ij},0}\left[-\left(J_{ij}^{*}\right)^{2}/\left(2\epsilon\right)-\ln\sqrt{2\pi\epsilon}\right]\\
	& +\delta_{\phi_{ij},1}\sum_{\varsigma=\pm1}\delta_{z_{j},\varsigma}\mathbbm{1}_{\varsigma J_{ij}^*>0}\left[-\left(\ln\zeta J_{ij}^{*}-\mu_{J}^{\zeta}\right)^{2}/\left(2v_{J}^{\zeta}\right)-\ln\sqrt{2\pi v_{J}^{\zeta}}\right]\Bigg\},
\end{align*}
and 
\begin{align*}
	& \sum_{\boldsymbol{\phi}}\sum_{\boldsymbol{z}}P_{\text{old}}\left(\boldsymbol{\phi}\left|\boldsymbol{z}\right.\right)P_{\text{old}}\left(\boldsymbol{z}\right)\ln p\left(\boldsymbol{\phi}\left|\boldsymbol{z},\boldsymbol{\rho}\right.\right)= \sum_{j}\sum_{\boldsymbol{\phi}}\sum_{\boldsymbol{z}}P_{\text{old}}\left(\boldsymbol{\phi}\left|\boldsymbol{z}\right.\right)P_{\text{old}}\left(\boldsymbol{z}\right)\ln p\left(\left\{ \phi_{kj}\right\} _{k}\left|z_{j},\boldsymbol{\rho}\right.\right)\\
	= & \sum_{j}\sum_{\boldsymbol{\phi}}\sum_{\boldsymbol{z}}\left\{ \prod_{a\neq j}\left[P_{\text{old}}\left(\left\{ \phi_{ka}\right\} _{k}\left|z_{a}\right.\right)P_{\text{old}}\left(z_{a}\right)\right]\right\} P_{\text{old}}\left(\left\{ \phi_{kj}\right\} _{k}\left|z_{j}\right.\right)P_{\text{old}}\left(z_{j}\right)\ln p\left(\left\{ \phi_{kj}\right\} _{k}\left|z_{j},\boldsymbol{\rho}\right.\right)\\
	= & \sum_{j}\sum_{\left\{ \phi_{kj}\right\} _{k}}\sum_{z_{j}=\pm1}P_{\text{old}}\left(\left\{ \phi_{kj}\right\} _{k}\left|z_{j}\right.\right)P_{\text{old}}\left(z_{j}\right)\ln p\left(\left\{ \phi_{kj}\right\} _{k}\left|z_{j},\boldsymbol{\rho}\right.\right)\\
	= & \sum_{j}\sum_{i}\sum_{z_{j}=\pm1}P_{\text{old}}\left(z_{j}\right)\sum_{\left\{ \phi_{kj}\right\} _{k}}\left[\prod_{b\neq i}P_{\text{old}}\left(\phi_{bj}\left|z_{j}\right.\right)\right]P_{\text{old}}\left(\phi_{ij}\left|z_{j}\right.\right)\ln p\left(\phi_{ij}\left|z_{j},\boldsymbol{\rho}\right.\right)\\
	= & \sum_{j}\sum_{i}\sum_{z_{j}=\pm1}P_{\text{old}}\left(z_{j}\right)\sum_{\phi_{ij}=1,0}P_{\text{old}}\left(\phi_{ij}\left|z_{j}\right.\right)\ln p\left(\phi_{ij}\left|z_{j},\boldsymbol{\rho}\right.\right)\\
	= & \sum_{i}\sum_{j}\sum_{z_{j}=\pm1}\sum_{\phi_{ij}=1,0}P_{\text{old}}\left(\phi_{ij}\left|z_{j}\right.\right)P_{\text{old}}\left(z_{j}\right)\ln p\left(\phi_{ij}\left|z_{j},\boldsymbol{\rho}\right.\right)\\
	= & \sum_{i}\sum_{j}\sum_{z_{j}=\pm1}\sum_{\phi_{ij}=1,0}P_{\text{old}}\left(\phi_{ij}\left|z_{j}\right.\right)P_{\text{old}}\left(z_{j}\right)\left[\delta_{\phi_{ij},1}\left(\ln\theta_{ij}-al_{ij}\right)+\delta_{\phi_{ij},0}\ln\left(1-\theta_{ij}e^{-al_{ij}}\right)\right],
\end{align*}
and 
\begin{align*}
	& \sum_{\boldsymbol{z}}P_{\text{old}}\left(\boldsymbol{z}\right)\ln p\left(\boldsymbol{z}\left|\boldsymbol{\rho}\right.\right)\\
	= & \sum_{i}\sum_{\boldsymbol{z}}\left[\prod_{a\neq i}P_{\text{old}}\left(z_{a}\right)\right]P_{\text{old}}\left(z_{i}\right)\ln p\left(z_{i}\left|\boldsymbol{\rho}\right.\right)\\
	= & \sum_{i}\sum_{z_{i}=\pm1}P_{\text{old}}\left(z_{i}\right)\ln p\left(z_{i}\left|\boldsymbol{\rho}\right.\right)\\
	= & \sum_{i}\sum_{z_{i}=\pm1}P_{\text{old}}\left(z_{i}\right)\left[\delta_{z_{i},+1}\ln\gamma+\delta_{z_{i},-1}\ln\left(1-\gamma\right)\right].
\end{align*}
We remark that the posterior probability of $\phi_{ij}$ depends on $z_j$ due to the dependence of $\phi_{ij}$ on $J_{ij}$, which depends on $z_j$. Therefore, the expected complete data log likelihood simplifies to 
\begin{alignb}
	\mathcal{Q}&=\ln P\left(\boldsymbol{s}\left|\boldsymbol{J}^{*},\boldsymbol{H}^{*}\right.\right)+\ln p\left(\boldsymbol{H}^{*}\left|\boldsymbol{\rho}\right.\right)+\sum_{\boldsymbol{\phi}}\sum_{\boldsymbol{z}}P\left(\boldsymbol{\phi},\boldsymbol{z}\left|\boldsymbol{s},\boldsymbol{J}^{*},\boldsymbol{H}^{*},\boldsymbol{\rho}^{\text{old}}\right.\right)\ln p\left(\boldsymbol{J}^{*},\boldsymbol{\phi},\boldsymbol{z}\left|\boldsymbol{\rho}\right.\right)\\
    &=\sum_{t}\sum_{i}\left\{ \left(H_{i}^{*}+\sum_{j}J_{ij}^{*}s_{j}^{t-1}\right)s_{i}^{t}-\ln\left[2\cosh\left(H_{i}^{*}+\sum_{j}J_{ij}^{*}s_{j}^{t-1}\right)\right]\right\} \\
    &\qquad+\sum_{i}\left\{ \frac{-\left(H_{i}^{*}-\mu_{H}\right)^{2}}{2v_{H}}-\ln\sqrt{2\pi v_{H}}\right\} \\
    &\qquad+\sum_{i}\sum_{j}\sum_{z_{j}=\pm1}\sum_{\phi_{ij}=1,0}P_{\text{old}}\left(\phi_{ij}\left|z_{j}\right.\right)P_{\text{old}}\left(z_{j}\right)\Bigg\{\delta_{\phi_{ij},0}\left[-\left(J_{ij}^{*}\right)^{2}/\left(2\epsilon\right)-\ln\sqrt{2\pi\epsilon}\right]\\
    &\qquad\qquad+\delta_{\phi_{ij},1}\sum_{\varsigma=\pm1}\delta_{z_{j},\varsigma}\mathbbm{1}_{\varsigma J_{ij}^*>0}\left[-\left(\ln\zeta J_{ij}^{*}-\mu_{J}^{\zeta}\right)^{2}/\left(2v_{J}^{\zeta}\right)-\ln\sqrt{2\pi v_{J}^{\zeta}}\right]\Bigg\}\\
    &\qquad+\sum_{i}\sum_{j}\sum_{z_{j}=\pm1}\sum_{\phi_{ij}=1,0}P_{\text{old}}\left(\phi_{ij}\left|z_{j}\right.\right)P_{\text{old}}\left(z_{j}\right)\left[\delta_{\phi_{ij},1}\left(\ln\theta_{ij}-al_{ij}\right)+\delta_{\phi_{ij},0}\ln\left(1-\theta_{ij}e^{-al_{ij}}\right)\right]\\
    &\qquad+\sum_{i}\sum_{z_{i}=\pm1}P_{\text{old}}\left(z_{i}\right)\left[\delta_{z_{i},+1}\ln\gamma+\delta_{z_{i},-1}\ln\left(1-\gamma\right)\right],
\end{alignb}
allowing us to perform the micro EM more efficiently.

\subsubsection{Micro E-step}

In the micro E-step, we evaluate the posterior probabilities $P\left(\boldsymbol{\phi}\left|\boldsymbol{z},\boldsymbol{s},\boldsymbol{J}^{*},\boldsymbol{H}^{*},\boldsymbol{\rho}^{\text{old}}\right.\right)$
and $P\left(\boldsymbol{z}\left|\boldsymbol{s},\boldsymbol{J}^{*},\boldsymbol{H}^{*},\boldsymbol{\rho}^{\text{old}}\right.\right)$
for $\boldsymbol{\phi}$ and $\boldsymbol{z}$, conditioned on $\boldsymbol{\rho}^{\text{old}}$ evaluated in the previous micro M-step as well as $\boldsymbol{J}^{*}$ and $\boldsymbol{H}^{*}$ obtained in the last macro E-step. In particular,
\begin{align}
	P\left(\boldsymbol{\phi}\left|\boldsymbol{z},\boldsymbol{s},\boldsymbol{J}^{*},\boldsymbol{H}^{*},\boldsymbol{\rho}^{\text{old}}\right.\right) & =\prod_{ij}P\left(\phi_{ij}\left|\boldsymbol{z},\boldsymbol{s},\boldsymbol{J}^{*},\boldsymbol{H}^{*},\boldsymbol{\rho}^{\text{old}}\right.\right),\text{ where}\\
	P\left(\phi_{ij}\left|\boldsymbol{z},\boldsymbol{s},\boldsymbol{J}^{*},\boldsymbol{H}^{*},\boldsymbol{\rho}^{\text{old}}\right.\right) & =p\left(\phi_{ij}\left|z_{j},J_{ij}^{*},\boldsymbol{\rho}^{\text{old}}\right.\right) \nonumber\\
	& =\frac{p\left(J_{ij}^{*}\left|\phi_{ij},z_{j},\boldsymbol{\rho}^{\text{old}}\right.\right)p\left(\phi_{ij}\left|z_{j},\boldsymbol{\rho}^{\text{old}}\right.\right)}{\sum_{\phi_{ij}'}p\left(J_{ij}^{*}\left|\phi_{ij}',z_{j},\boldsymbol{\rho}^{\text{old}}\right.\right)p\left(\phi_{ij}'\left|z_{j},\boldsymbol{\rho}^{\text{old}}\right.\right)};
\end{align}
and 
\begin{align}
	P\left(\boldsymbol{z}\left|\boldsymbol{s},\boldsymbol{J}^{*},\boldsymbol{H}^{*},\boldsymbol{\rho}^{\text{old}}\right.\right) & =\prod_{j}P\left(z_{j}\left|\boldsymbol{s},\boldsymbol{J}^{*},\boldsymbol{H}^{*},\boldsymbol{\rho}^{\text{old}}\right.\right),\text{ where} \\
	P\left(z_{j}\left|\boldsymbol{s},\boldsymbol{J}^{*},\boldsymbol{H}^{*},\boldsymbol{\rho}^{\text{old}}\right.\right) & =p\left(z_{j}\left|\left\{ J_{ij}^{*}\right\} _{i},\boldsymbol{\rho}^{\text{old}}\right.\right) \nonumber \\
	& =\frac{p\left(\left\{ J_{ij}^{*}\right\} _{i}\left|z_{j},\boldsymbol{\rho}^{\text{old}}\right.\right)p\left(z_{j}\left|\boldsymbol{\rho}^{\text{old}}\right.\right)}{\sum_{z_{j}'}\left[p\left(\left\{ J_{ij}^{*}\right\} _{i}\left|z_{j}',\boldsymbol{\rho}^{\text{old}}\right.\right)p\left(z_{j}'\left|\boldsymbol{\rho}^{\text{old}}\right.\right)\right]} \nonumber \\
	& =\frac{p\left(z_{j}\left|\boldsymbol{\rho}^{\text{old}}\right.\right)\prod_{i}\left[\sum_{\phi_{ij}'=1,0}p\left(J_{ij}^{*}\left|\phi_{ij}',z_{j},\boldsymbol{\rho}^{\text{old}}\right.\right)p\left(\phi_{ij}'\left|z_{j},\boldsymbol{\rho}^{\text{old}}\right.\right)\right]}{\sum_{z_{j}'}p\left(z_{j}'\left|\boldsymbol{\rho}^{\text{old}}\right.\right)\prod_{i}\left[\sum_{\phi_{ij}'=1,0}p\left(J_{ij}^{*}\left|\phi_{ij}',z_{j}',\boldsymbol{\rho}^{\text{old}}\right.\right)p\left(\phi_{ij}'\left|z_{j}',\boldsymbol{\rho}^{\text{old}}\right.\right)\right]}.
\end{align}
The updated posterior probabilities are then passed to the micro M-step for evaluating the updated $\boldsymbol{\rho}$.

\subsubsection{Micro M-step}

In the micro M-step, we use the posterior probabilities $P\left(\boldsymbol{\phi}\left|\boldsymbol{z},\boldsymbol{s},\boldsymbol{J}^{*},\boldsymbol{H}^{*},\boldsymbol{\rho}^{\text{old}}\right.\right)$
and $P\left(\boldsymbol{z}\left|\boldsymbol{s},\boldsymbol{J}^{*},\boldsymbol{H}^{*},\boldsymbol{\rho}^{\text{old}}\right.\right)$,
as well as the $\boldsymbol{J}^{*}$ and $\boldsymbol{H}^{*}$ obtained in the last macro E-step, to update the values of $\boldsymbol{\rho}$. The derivatives of $\mathcal{Q}$ with respect to all variables in $\boldsymbol{\rho}$ except $a$ give:
\begin{align}
	\frac{\partial\mathcal{Q}}{\partial\gamma} & =\sum_{i}\sum_{z_{i}=\pm1}P_{\text{old}}\left(z_{i}\right)\left[\delta_{z_{i},+1}\frac{1}{\gamma}+\delta_{z_{i},-1}\frac{1}{1-\gamma}\right],\\
	\frac{\partial\mathcal{Q}}{\partial\mu_{J}^{\zeta}} & =\sum_{i}\sum_{j}P_{\text{old}}\left(\phi_{ij}=1\left|z_{j}=\zeta\right.\right)P_{\text{old}}\left(z_{j}=\zeta\right)\mathbbm{1}_{\varsigma J_{ij}^*>0}\frac{\ln\zeta J_{ij}^{*}-\mu_{J}^{\zeta}}{v_{J}^{\zeta}},\text{ for }\zeta=\pm1\\
	\frac{\partial\mathcal{Q}}{\partial v_{J}^{\zeta}} & =\sum_{i}\sum_{j}P_{\text{old}}\left(\phi_{ij}=1\left|z_{j}=\zeta\right.\right)P_{\text{old}}\left(z_{j}=\zeta\right)\mathbbm{1}_{\varsigma J_{ij}^*>0}\frac{\left(\ln\zeta J_{ij}^{*}-\mu_{J}^{\zeta}\right)^{2}-v_{j}^{\zeta}}{2\left(v_{J}^{\zeta}\right)^{2}},\text{ for }\zeta=\pm1\\
	\frac{\partial\mathcal{Q}}{\partial\mu_{H}} & =\sum_{i}\left[\frac{H_{i}^*-\mu_{H}}{v_{H}}\right],\\
	\frac{\partial\mathcal{Q}}{\partial\mu_{H}} & =\sum_{i}\left[\frac{\left(H_{i}^*-\mu_{H}\right)^{2}-v_{H}}{2v_{H}^{2}}\right].
\end{align}
By setting them to zero, we obtain
\begin{align}
	\gamma & =\frac{\sum_{i}P_{\text{old}}\left(z_{i}=+1\right)}{N}\\
	\mu_{J}^{\zeta} & =\frac{\sum_{i}\sum_{j}P_{\text{old}}\left(\phi_{ij}=1\left|z_{j}=\zeta\right.\right)P_{\text{old}}\left(z_{j}=\zeta\right)\mathbbm{1}_{\varsigma J_{ij}^*>0}\ln\zeta J_{ij}^{*}}{\sum_{i}\sum_{j}P_{\text{old}}\left(\phi_{ij}=1\left|z_{j}=\zeta\right.\right)P_{\text{old}}\left(z_{j}=\zeta\right)\mathbbm{1}_{\varsigma J_{ij}^*>0}},\text{ for }\zeta=\pm1 \label{eq_mu_J}\\
	v_{J}^{\zeta} & =\frac{\sum_{i}\sum_{j}P_{\text{old}}\left(\phi_{ij}=1\left|z_{j}=\zeta\right.\right)P_{\text{old}}\left(z_{j}=\zeta\right)\mathbbm{1}_{\varsigma J_{ij}^*>0}\left(\ln\zeta J_{ij}^{*}-\mu_{J}^{\zeta}\right)^{2}}{\sum_{i}\sum_{j}P_{\text{old}}\left(\phi_{ij}=1\left|z_{j}=\zeta\right.\right)P_{\text{old}}\left(z_{j}=\zeta\right)\mathbbm{1}_{\varsigma J_{ij}^*>0}},\text{ for }\zeta=\pm1 \label{eq_v_J}\\
	\mu_{H} & =\frac{\sum_{i}H_{i}^*}{N}, \label{eq_mu_H}\\
	v_{H} & =\frac{\sum_{i}\left(H_{i}^*-\mu_{H}\right)^{2}}{N}. \label{eq_v_H}
\end{align}

The micro E and M-steps are iterating alternatively until convergence, and the converged hyperparameters $\boldsymbol{\rho}^{\varXi}$, as well as the posterior probabilities $P\left(\boldsymbol{\phi}\left|\boldsymbol{z},\boldsymbol{s},\boldsymbol{J}^{*},\boldsymbol{H}^{*},\boldsymbol{\rho}^{\Xi}\right.\right)$ and $P\left(\boldsymbol{z}\left|\boldsymbol{s},\boldsymbol{J}^{*},\boldsymbol{H}^{*},\boldsymbol{\rho}^{\Xi}\right.\right)$ are then fixed to find the optimal $\boldsymbol{J}^{*}$ and $\boldsymbol{H}^{*}$ in the next macro E-step.

As an overview - we execute the macro E and M steps iteratively until convergence, and within each macro M-step, we evaluate the hyperparameters and the posterior probabilities by conducting the micro E-step and M-step iteratively until convergence. Afterwards, the structure parameters for neuronal nature $\boldsymbol{z}^{*}$ and link existence $\boldsymbol{\phi}^{*}$ are selected by choosing the highest probability states from the micro E-step , i.e. 
\begin{align}
	z_{j}^{*} & =\underset{z_{j}'}{\text{argmax}}P\left(z_{j}'\left|\boldsymbol{s},\boldsymbol{J}^{*},\boldsymbol{H}^{*},\boldsymbol{\rho}^{\Xi}\right.\right),\text{ and }\\
	\phi_{ij}^{*} & =\underset{\phi_{ij}'}{\text{argmax}}P\left(\phi_{ij}'\left|\boldsymbol{s},\boldsymbol{J}^{*},\boldsymbol{H}^{*},\boldsymbol{\rho}^{\Xi}\right.\right) \nonumber \\
	& =\underset{\phi_{ij}'}{\text{argmax}}\sum_{z_{j}'}P\left(\phi_{ij}'\left|z_{j}',\boldsymbol{s},\boldsymbol{J}^{*},\boldsymbol{H}^{*},\boldsymbol{\rho}^{\Xi}\right.\right)P\left(z_{j}'\left|\boldsymbol{s},\boldsymbol{J}^{*},\boldsymbol{H}^{*},\boldsymbol{\rho}^{\Xi}\right.\right). 
\end{align}

\section{Additional Results}
To validate the performance of our method on benchmark data, we genereated additional data sets from synthetic and in-silico models. In particular, to show our inference method performs well on biological neuronal systems compared to the existing methods, we test our approach with synthetic data. 

\subsection{Kinetic Ising model}
To examine the efficacy of our inference algorithm and compare its performance to that of existing approaches, we first use synthetic data generated directly from a kinetic Ising model~\cite{mezard2011exact}, a similar model to the celebrated Ising model but where interactions between spins are non-symmetric. In particular, we generate data from a kinetic Ising model with system size $N=50$. To simulate biological systems, we assigned $80\%$ ($20\%$) of the neurons (or spins) to be excitatory (inhibitory), where $J_{ij} \geq 0$ ($\leq 0$) for all $i$ if $j$ is an excitatory (inhibitory) neuron. Additionally, the probability of two spins to be connected decays exponentially with the distance between them. Synthetic data is generated using Monte-Carlo simulation for $3000$ steps. We then infer the structure variables $\boldsymbol{J}$ given the synthetically-generated Monte-Carlo data. Subsequently, we predict the structure variables (connectivity, neuronal type and synaptic strengths) using the inferred structure and the known underlying variable values. The inference method also requires one to evaluate the equal-time covariance matrix $\boldsymbol{C}=\left\{ \left\langle s_{i}^{t}s_{j}^{t}\right\rangle _{t}\right\} _{i,j}$ and the delayed time covariance matrix $\boldsymbol{D}=\left\{ \left\langle s_{i}^{t}s_{j}^{t-1}\right\rangle _{t}\right\} _{i,j}$; we also use these matrices, of both inferred and true data, to measure the accuracy of the inference method.
\begin{figure}[h]
	\centering
	\includegraphics[width=0.65\linewidth, trim=0 0 0 0, clip]{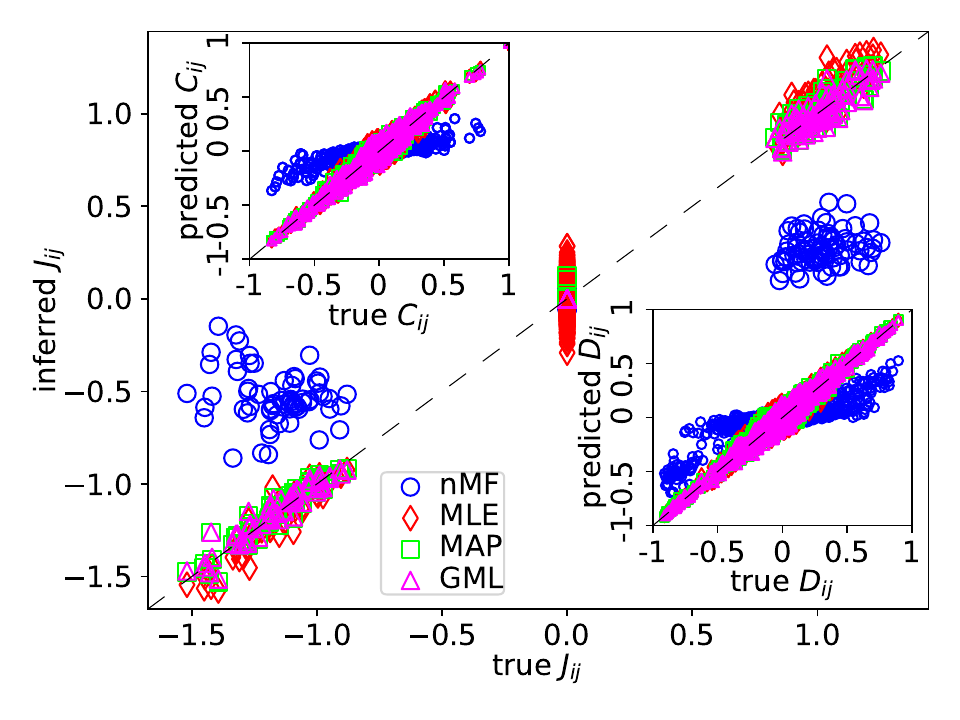} 
	\caption{Inferred coupling strength $\boldsymbol{J}$ obtained using  nMf, MLE, MAP and GML compared to the true coupling strengths generated by a synthetic model. Top left inset: the predicted equal time covariance matrix values $\boldsymbol{C}$ against the true values. Bottom right inset: the predicted delayed time covariance matrix values $\boldsymbol{D}$ against the true values. Synthetic data are generated from a realization of kinetic Ising model simulation with $N$=50, the ratio between positive and negative $\boldsymbol{J}$ is 1:4, mean values of positive and negative $\boldsymbol{J}$ are 1 and -1.35 respectively, the fraction of effective links is $6\%$, the time steps of the simulation is $3000$.}
	\label{fig_syn}
\end{figure}

We compare our GML approach against the naive mean field (nMF), maximum likelihood estimation (MLE) and maximum a posteriori (MAP) approach, by plotting the inferred values of $J_{ij}$ against the true value in~\fig{fig_syn}. Briefly, nMF refers to the simple factorization of the joint probability of all relevant variables into individual probabilities, MLE refers to maximizing the probability of the data given model parameters, MAP to maximizing the probability of parameter values given data and the GML to an iterative process whereby model hyper-parameters and variable values are estimated and optimized recursively.

\fig{fig_syn} shows the accuracy of the inferred synaptic strengths $\boldsymbol{J}$, predicted equal-time covariance matrix $\boldsymbol{C}$ and delayed time covariance matrix $\boldsymbol{D}$; the better the scatter plot aligns with the dotted line $y=x$, the more accurate is the structure parameters' inference and its alignment with the data. We observe that nMF results are inferior in performance compared to the other methods; naive mean-field works best when $J_{ij}$ are Gaussian distributed with zero mean and small variance, while in the more realistic case we simulate $J_{ij}$ follow a mixture of distributions that are far from zero. The predicted values using MLE already align very well with the true values, but many disconnected neuron-pairs $J_{ij}=0$ are inferred to have nonzero values, due to the fact that link existence is not considered separately in MLE. 

To address the issue of  non-existing links we employ the EM algorithm. Remarkably, even after a single macro EM step (jointly with MAP), except two true zero links, all true zero links are inferred as zero (disconnected). Moreover, if we employ the full GML procedure until convergence, we can see that neuron type classification (excitatory/inhibitory/zero) for all $J_{ij}$ is completely correct. This suggests that our algorithm excels at identifying effective linkages between neurons (spins). In addition, we can see that the error of inference using the GML is significantly lower than those achieved by other methods. The predicted $\boldsymbol{C}$ and $\boldsymbol{D}$ covariance matrices are plotted against the true values as shown in the insets of~\fig{fig_syn}. The prediction accuracy for $\boldsymbol{C}$ and $\boldsymbol{D}$ are similar for MLE, MAP and GML. These results suggest that our model and inference algorithm are appropriate for structural inference on biological neural network, especially when $J_{ij}$ follows a mixture of distributions.

\subsection{Experimental data - In-silico homogeneous network}
In the main text we studied the performance of our algorithm on emulated in-silico data over patterned substrates and showed that our method performs very well on neuronal type classification and link existence. Noted that patterned substrates restrict neurons from different modules connecting to each other, thus might reduce the solution space of $\boldsymbol{J}$, $\boldsymbol{H}$, $\boldsymbol{z}$ and $\boldsymbol{\phi}$, making the inference problem easier to solve compared to homogeneous setup. Therefore, one would expect better inference results to be obtained from the activities data generated from the patterned substrates model. Here we test our algorithm on emulated in-silico data with homogeneous network structure, to verify the performance of accuracy of neuronal type classification and link existence on non-structured substrates. 

Homogeneous neuronal network with $156$ neurons is generated using the same emulator as the in-silico model with patterening substrates studied in the main text. Next, spontaneous neuronal activity data is generated and used as the input for the effective structure inference using MLE, MAP and the GML. For the prior distributions, based on the statistical information already known from the in-silico model, we fix $\gamma=0.8$, $a=0.1$ and $\theta_{ij}=0.9$ for all neurons $i$ and $j$. The values of the other hyperparameters, that are used for the both MAP inference and the inintal values of GML, are evaluated using Eq.~[\ref{eq_mu_J}-\ref{eq_v_H}], given the $\boldsymbol{J}$ and $\boldsymbol{H}$ inferred by MLE. 

\begin{figure*}[h]
	\centering
	\begin{tabular}{@{} c @{} c @{} c @{}}
		\begin{tabular}{l}
			\vspace{-0.6em}A \\
			\includegraphics[width=0.35\linewidth, trim=0 0 0 0, clip]{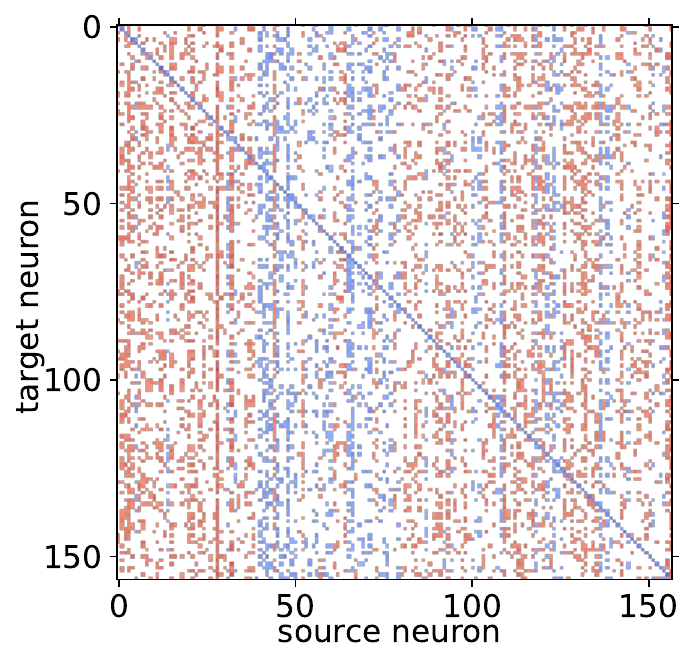} \\
		\end{tabular}
		&
		\hspace{-1em}	\begin{tabular}{l}
			B \\
			\includegraphics[width=0.32\linewidth, trim=0 0 0 0, clip]{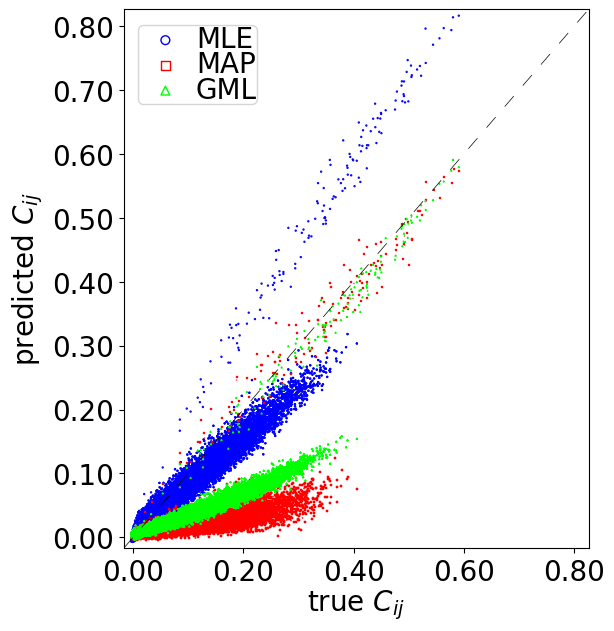} \\
		\end{tabular}
		&
		\hspace{-1em}
		\begin{tabular}{l}
			C \\
			\includegraphics[width=0.32\linewidth, trim=0 0 0 0, clip]{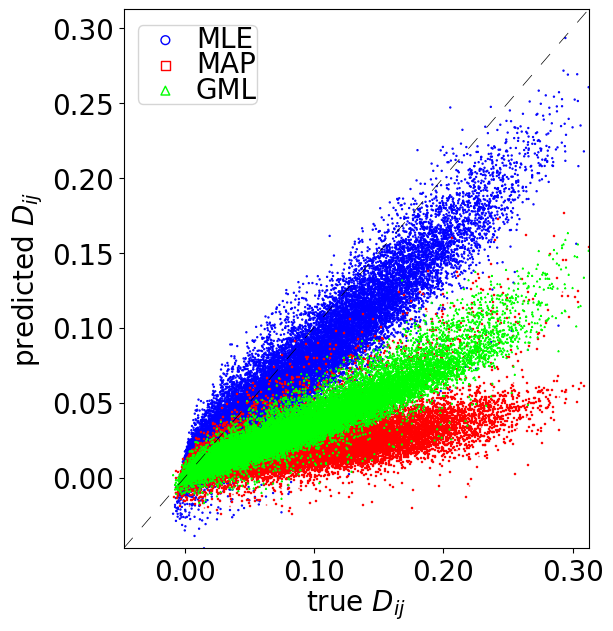} 
		\end{tabular}
	\end{tabular}
	\caption{
		A: The inferred structure $\boldsymbol{J}$ of the in-silico homogeneous model obtained using GML and represented by a connectivity matrix. Each entry corresponds to the coupling strength $J_{ij}$. A positive (negative) strength colored in red (blue) refers to excitatory (inhibitory) signals sent from source ($i$) to target ($j$) neurons. B and C: The predicted equal-time covariance $\boldsymbol{C}$ and delayed time covariance $\boldsymbol{D}$ against the true values evaluated from data, respectively.
	}
	\label{fig_si_silico}
\end{figure*}
\begin{table}[h]
	\begin{center}
		\scalebox{1}{
			\begin{tabular}{||c| c|c| c ||} 
				\hline
				\textbf{Measure }& \textbf{ \red{TPR}/\blue{TNR} } & \textbf{\red{PPV}/\blue{NPV}}& \textbf{Random}\\ [0.5ex] 
				\hline\hline
				\textbf{Excitatory} & \red{0.81} & \red{\textbf{0.94}}  & 0.8\\ [0.5ex] 
				\hline
				\textbf{Inhibitory} & \blue{0.81}  & \blue{\textbf{0.52}} & 0.2 \\ 
				\hline
				\textbf{Overall}& n/a &   \textbf{0.81} & 0.68  \\[0.5ex] 
				\hline
			\end{tabular}
		}
	\end{center}
	\caption{Success measures in identifying neuron type: true positive rate (TPR -  sensitivity), True negative rate  (TNR -  specificity) and positive predictive value (PPV) of our in-silico homogeneous model study.}
	\label{tab_si_neuronal_nature}
\end{table}
The connectivity matrix showing the inferred coupling strengths $\boldsymbol{J}$ using GML is shown in \fig{fig_si_silico}A, where inhibitory and excitatory connections are colored in blue and red respectively. Similar to the other results we obtained in-silico and in-vitro experiments, we can see that the diagonal connections are mostly inhibitory, which may reflects the fact that neurons are less likely to have consecutive firing due to the reduced potential. To measure the precision of neuronal type prediction, the positive predictive value (PPV), negative predictive value (NPV) of $P\left(z_{\mbox{true}}\left|z_{\mbox{inferred}}\right.\right)$, as well as the true positive rate (TPR) and the true negative rate (TNR) showing $P\left(z_{\mbox{inferred}} = +/- \left| z_{\mbox{true}} = +/- \right.\right)$, are shown in \tab{tab_si_neuronal_nature}, where excitatory and inhibitory nature are the positive case and negative case respectively. The overall predictability is the weighted sum of the probabilities in both cases. We can see that the neuronal type prediction of our method outperform the prior-based random guess approach, while the predictability of the patterned substrates case is slightly higher only. The high predictability suggests that, our method still perform very well even on the homogeneous case.

\begin{figure}[h]
	\centering
	\includegraphics[width=0.50\linewidth, trim=0 0 0 0, clip]{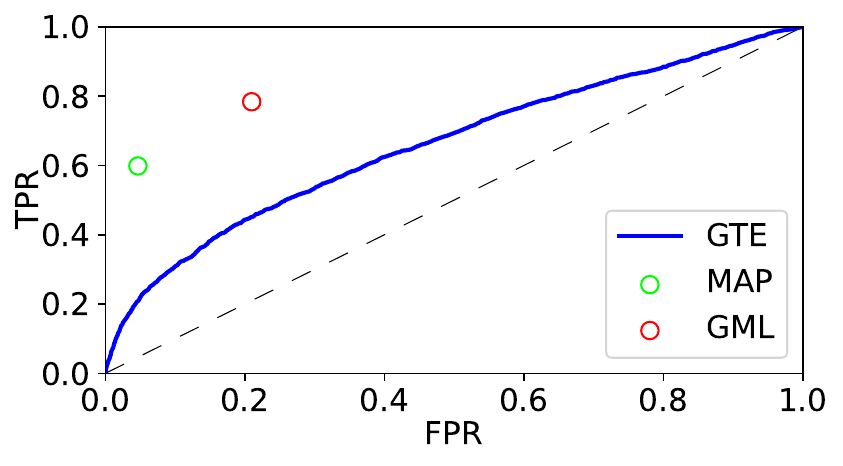} \\
	\caption{
		The receiver operating characteristic (ROC) curve plotting the true positive rate (TPR) against the false positive rate (FPR) of identifying effective links for in-silico experiments using generalized transfer entropy (GTE). The TPR and TFR using our MAP and GML are marked as green and red nodes respectively.
	}
	\label{fig_silico_GTE}
\end{figure}

Next, we test the performance of the link existence inference using MAP and GML, and compare them with the results obtained by generalized transfer entropy (GTE)~\cite{orlandi2014transfer}. The complete receiver operating characteristic (ROC) curve showing the TPR and FPR of identifying non-zero links using GTE with all threshold values is shown in \fig{fig_silico_GTE}. The corresponding TPR and FPR using GML are 78\% and 21\%, respectively, and 60\% and 4.7\% for MAP, respectively, are indicated by red and green circles in the figure as well. Similar to the main text, our method offers a higher TPR than GTE at the same FPR level, or a lower FPR at the same level of TPR. This suggest that our method performs well in identifying effective links even without the help of the patterned substrates. 

Similar to what we have done in the main text, we employ the inferred structure $\boldsymbol{J}$ and $\boldsymbol{H}$ using MLE, MAP and GML, to generate artificial neuronal activities through Monte Carlo simulation to validate how well the inferred structure describes the true model by comparing $\boldsymbol{C}$ and $\boldsymbol{D}$ with the true values, as shown in \fig{fig_si_silico}B and C respectively. We can see that in general, the predicted values closely align with the true values for all methods, for both $\boldsymbol{C}$ and $\boldsymbol{D}$. First of all, we can see that MLE has the best performance as anticipated, similar to that we observe in the pattering substrates system. This is because there are no restrictions and prior knowledge provided to the inference algorithm. Our GML algorithm performs very well in activity predictions and is better than MAP, which is indeed the GML with one iteration. This suggests that GML as the full evidence-based approach can improve the inference of neuronal activities by optimizing the values of hyperparameters.

In general, by studying the homogeneous in-silico model, we show that GML  performs very well in identifying neuronal types and link existence inference as well as the neuronal activities prediction, even without the constraints imposed by patterned substrates.

\bibliographystyle{prsty}
\bibliography{net_infer_ref}

\end{document}